\documentstyle[prl,aps]{revtex}
\begin{document}

%%%%%%%%%%%%%
%%% \newcommand
\newcommand{\be}{\begin{equation}}
\newcommand{\ee}{\end{equation}}
\newcommand{\beqn}{\begin{eqnarray}}
\newcommand{\eeqn}{\end{eqnarray}}
\newcommand{\eq}{equation }
\newcommand{\eqs}{equations }
\newcommand{\dq}{{\dot q}}
\newcommand{\ddq}{{\ddot q}}
\newcommand{\NP}[1]{Nucl. Phys. {\bf #1} }
\newcommand{\PL}[1]{Phys. Lett. {\bf #1} }
\newcommand{\ZP}[1]{Z. Phys. {\bf #1} }
\newcommand{\PR}[1]{Phys. Rev. {\bf #1} }
\newcommand{\PRL}[1]{Phys. Rev. Lett. {\bf #1} }
\newcommand{\JCP}[1]{J. Chem. Phys. {\bf #1} }
\newcommand{\JPG}[1]{J. Phys. {\bf G}: Nucl. Part. Phys. {\bf #1} }
\newcommand{\AP}[1]{Ann. Phys. {\bf #1} }
\newcommand{\RPP}[1]{Rep. on Prog. in Phys. {\bf #1} }
\newcommand{\ARNPS}[1]{Annu. Rev. Nucl. Part. Sci. {\bf #1} } 
\newcommand{\RMP}[1]{Rev. Mod. Phys. {\bf #1} } 
%
%%% End of the commands
%%%%%%%%%%%%%%%%%%%%%%%%%%%%%%%%%%

\title{Diffusion over a saddle with a Langevin equation
\footnote{to appear in Phys. Rev. E61}}
\author{Yasuhisa Abe$^{(a)}$, David Boilley$^{(a,b)}$, Bertrand G. 
Giraud$^{(c)}$ and Takahiro Wada$^{(d)}$}
\date{\today}  
\maketitle

\vspace{-27mm}
\rightline{YITP-99-71}

\vspace{30mm}
\leftline{\small $^{(a)}$ Yukawa Institute for Theoretical Physics, Kyoto
University, Kyoto 606-8502, Japan}
\leftline{\small $^{(b)}$ GANIL, BP 5027, 14 076 Caen Cedex 05, France}
\leftline{\small $^{(c)}$ Service de Physique Th\'eorique, DSM, CE Saclay, 
F-91191 Gif/Yvette, France}
\leftline{\small $^{(d)}$ Konan University, Okamoto 8-9-1, Higashinada, Kobe
658, Japan}

\vspace{1cm}

\begin{abstract}
The diffusion problem over a saddle is studied using a multi-dimensional
Langevin equation. An analytical solution is derived for a quadratic 
potential and the probability to pass over the barrier deduced. A very 
simple solution is given for the one dimension problem and a general 
scheme is shown for higher dimensions. 
\end{abstract}

%%%%%%%%%%%%%%%%%%%%%%%%%%%% Introduction

\section{Introduction}

The Langevin equation \cite{Lan} has been applied to most fields of physics. 
It was solved several times for parabolic potential wells, see e.g. 
\cite{Uhl,Chan}. As analytical solutions can be derived for quadratic 
potentials only, it has really rather become widely used in numerical 
simulations.

Our purpose is to establish an analytical expression for the diffusion 
over a potential barrier. In order to have a solvable problem, we assume 
that, around the saddle point, the potential can be approximated by 
quadratic functions. Some very simple expressions are obtained in one 
dimension. Because many processes obviously involve more than one 
coordinate, we extend our analysis to multi-dimensional cases. We thus 
derive an analytic expression for the distribution function of the 
Langevin equation, valid for multi-dimensional models, and then study the 
probability to over-pass the barrier.

Our approach is only valid for classical diffusion satisfying the 
dissipation-fluctuation theorem. A solution for the one dimension Langevin 
equation in the overdamped limit is derived in Ref. \cite{Jes} for L\'evy 
flights \cite{Levy}, but it cannot be simply extended to multi-dimensional 
Langevin equations.

The interest of our approach will be shown in the case of heavy-ion fusion 
problems, for which Langevin equation type simulations have been used by 
several groups \cite{Ari,Agu,Maz}. The very small cross-section of such a 
mechanism makes numerical simulations very difficult, because very large 
statistics have to be computed. Our analytical expressions, though using 
somewhat crude approximations, could be useful to extract some general 
trends. The problematics of realistic calculations is out of the scope of 
this paper, where we only discuss general considerations concerning the 
Langevin formalism.

%%%%%%%%%%%%%%%%%%%%%%%%%%%% Langevin Equation

\section{The Langevin equation}
\subsection{Introduction}

To study the diffusion over a 1-D parabolic potential barrier, $V(q_1)=
-m_1\omega_1^2 q_1^2/2$ with a given initial condition $q_{10}<0$ and 
$p_{10}>0$, the Langevin equation reads,
\be
\ddq_1 + \beta_1 \dq_1 -\omega_1^2 q_1 = r_1(t),
\label{Lang}
\ee
where $r_1(t)$ is a Gaussian stochastic force. As discussed in Appendix B, we 
rule out any anomalous diffusion process. The first two moments of this 
force are,
\be 
<r_1(t)> = 0 \qquad {\rm and } \qquad <r_1(t)r_1(t')>= \frac{2 T\beta_1}{m_1}
\, \delta(t-t'),
\ee
in agreement with the dissipation-fluctuation theorem. In the previous
equations, $T,$ $m_1$ and $\beta_1$ denote the temperature, the mass and the 
reduced friction, respectively. All these parameters are assumed to be
time, position and velocity independent, or at most very slowly varying, in 
the vicinity of the saddle. The symbol $<\,>$ indicates an ensemble average.

For any $n$-dimensional problem, one can generalize the previous 
approach, replacing the $(q_1,\dot q_1)$ variables by vectors $(Q,\dot Q),$
\be
\ddot Q + \beta \dot Q - \Omega^2 Q =  R(t) 
\, .
\label{canoni}
\ee
Such a canonical form of the problem results from two successive, very 
standard manipulations:
\par \noindent
- {\it i)} change the representation by transforming all tensors 
${\cal T},$ such as the friction tensor and the spring tensor, into 
a form $M^{-1/2} {\cal T} M^{-1/2},$ where $M$ is the usual mass tensor, 
naturally,
\par \noindent
- {\it ii)} change again the representation so that the spring tensor 
becomes diagonal.

More precisely, assume some initial representation with a vector of 
degrees of freedom $Z\equiv\{z_1,...z_n\}$, driven by a constant,
symmetric mass tensor $M,$ a constant, symmetric friction tensor 
${\cal G,}$ a constant, symmetric spring tensor ${\cal S}$ and a random 
vector force ${\cal F}.$ The initial dynamical equation reads,
\be
M \ddot Z + {\cal G} \dot Z - {\cal S} Z =  {\cal F}(t) 
\, .
\label{brute}
\ee
This is equivalent to,
\be
M^{1/2} \ddot Z + M^{-1/2}{\cal G} M^{-1/2} M^{1/2}\dot Z - 
M^{-1/2}{\cal S} M^{-1/2} M^{1/2} Z = M^{-1/2} {\cal F}(t) 
\, .
\label{demi}
\ee
Let now $U$ be that orthogonal matrix which lists the right eigenvectors 
of $M^{-1/2} {\cal S} M^{-1/2}$ as columns. Accordingly, 
$M^{-1/2} {\cal S} M^{-1/2} = U \Omega^2 U^{-1},$ where $\Omega^2$ is
diagonal. (Throughout this argument we rule out, naturally, those very 
exceptional cases where diagonalizations and/or inversions are singular.) 
Then the dynamical equation reads as well,
\be
U^{-1} M^{1/2} \ddot Z + U^{-1} M^{-1/2}{\cal G} M^{-1/2} U \ U^{-1} M^{1/2}
\dot Z - \Omega^2\ U^{-1} M^{1/2} Z = U^{-1} M^{-1/2} {\cal F}(t) 
\, .
\ee
With the definitions, $Q \equiv U^{-1} M^{1/2} Z,$ 
$\beta \equiv U^{-1} M^{-1/2}{\cal G} M^{-1/2} U$ 
and $R(t) = U^{-1} M^{-1/2} {\cal F}(t),$ this is nothing but
the canonical form, Eq.(\ref{canoni}).

For 2-D for instance, we may obtain the first final degree of freedom 
as a ``valley'' direction to the saddle and the other final degree 
as a ``confining'' direction,
\be 
Q=\left[\matrix{ q_1 \cr q_2}\right], \qquad 
\Omega^2=\left[\matrix{ \omega_1^2 & 0 \cr 0 & -\omega_2^2}\right],  
\qquad  \qquad \beta = \left[\matrix{\beta_1 & \beta_{12} \cr \beta_{12} &
\beta_2}\right],
\ee
and $R(t)$ is a random force with two components. Such two components are 
correlated when $\beta$ is non-diagonal but can be related, through a 
suitable matrix $\Gamma,$ to a vector of independent random numbers,
\be 
R(t)= \Gamma \left[\matrix{\nu_1(t) \cr \nu_2(t)} \right],
\ee 
with
\be 
<\nu_1(t)\nu_1(t')>=\,  \delta(t-t'),\quad <\nu_2(t)\nu_2(t')>=
\,  \delta(t-t')  \quad {\rm and} \quad <\nu_1(t)\nu_2(t)>=0
\, .
\label{norma}
\ee
The matrix $\Gamma$ is real, but usually not symmetric. 
The fluctuation-dissipation theorem, incidentally, which is easy to
derive from the initial form, Eq.(\ref{brute}), states that
$\Gamma\Gamma^T=2T\beta$, where $T$ is the temperature and where the 
superscript $T$ indicates transposition. All necessary details are
found in Appendix A.

It will be stressed again here that the matrices $\beta,$ $\Omega^2$ and 
$\Gamma$ take into account an overall multiplication of motion equations 
by the inverse of the (square root of the) mass tensor. It is easy to 
prove that this manipulation does not change the signs of the eigenvalues 
of the resulting matrices, and that such resulting matrices are usually non
diagonal. In turn, the final spring tensor $\Omega^2$ can be made diagonal 
by the additional manipulation {\it ii)}. 

The generalization to $n$ dimensions is trivial.
 
\subsection{Analytical solution}

Defining $P=\dot Q$, 
it is easy to transform Eq.(\ref{canoni}) into a first-order differential 
equation, in a (2$n$)-D space,
\be
{d \over dt} \left[\matrix{Q \cr P}\right] = 
\left[\matrix{0 & 1 \cr \Omega^2 & -\beta}\right] 
\left[\matrix{ Q \cr P}\right] + \left[\matrix{0 \cr R} \right]
\, .
\label{grande}
\ee 
In the following we call ``drift matrix'' (DM) that block matrix 
${\cal D}=\left[\matrix{0 & 1 \cr \Omega^2 & -\beta}\right] $
appearing in Eq.(\ref{grande}). In the upper-left and upper-right corners 
of this $2n \times 2n$ matrix, the symbols 0 and 1 denote, respectively, 
the null and the unit $n \times n$ matrices, naturally. Whenever this DM
can be diagonalized, and in the limit where the time and space derivatives 
of all physical parameters such as masses, drifts, frequencies, etc. can 
be neglected, the previous system can be transformed into,
\be
{d \over dt} \left[\matrix{X_1 \cr \dot{.} \cr X_{2n}}\right] =
D \left[\matrix{X_1 \cr \dot{.} \cr X_{2n}}\right] + 
\theta^{-1}\left[\matrix{0 \cr R} \right], 
\ee
where $D$ is the diagonal drift matrix, 
\be
D =\theta^{-1}{\cal D}\theta = \left[\matrix{a_1 & 0 & 0 \cr 0 &..&0\cr 0&0& a_{2n}}\right] ,
\label{DDM}
\ee
and $\theta$ is the ``rotation'' matrix,
\be
\left[\matrix{q_1 \cr\dot{.} \cr q_n \cr p_1 \cr \dot{.}\cr p_n}\right] =
\theta\left[\matrix{X_1\cr\dot{.}\cr\dot{.}\cr\dot{.}\cr\dot{.} \cr 
X_{2n}}\right].
\ee 
This matrix $\theta$ is the matrix of right (column) eigenvectors of
the DM. Any normalization may be chosen for such eigenvectors.

It can be stressed here that the eigenmodes $X_i$ are linear
combinations of both positions and momenta, hence the saddle dynamics
should be visualized in phase space rather than coordinate space only.
All such eigenmodes are expected to decay exponentially with time,
except just one, corresponding to a resulting preferred direction along 
the valley, in phase space.

In that same limit where the derivatives of the physical parameters
(masses, etc.) can be neglected, such first order differential
equations can be formally integrated into,
\be
x_i(t) \equiv X_i e^{-a_it}-X_{io} = \int_0^t d\tau\, e^{-a_i\tau}\,
[\alpha_{i1} \nu_1(\tau)+ ... +\alpha_{in} \nu_n(\tau)],\ \ \
i=1,2,...,2n
\, .
\label{Eula}
\ee
Here the $\alpha_{ij}$'s are defined from the effects of both $\theta^{-1}$ 
and $\Gamma$ matrices (in matrix notation, 
$\alpha=\theta^{-1}\left[\matrix{0 \cr \Gamma \cr}\right],$ 
where $\alpha$ is a $2n \times n$ matrix, and $0$ and $\Gamma$ are 
$n \times n$ ones) and the $\nu_i$'s are uncorrelated white random numbers, 
extending Eqs.(\ref{norma}) to $n$ dimensions. Then, the Euler type variables 
$(x_1, ..., x_{2n})$ should have the same statistical properties as those of 
\be
\left( \int_0^t d\tau \, e^{-a_1\tau} [\alpha_{11} \nu_1(\tau)+
... +\alpha_{1n} \nu_n(\tau)],\ ...\ ,\int_0^t d\tau 
e^{-a_{2n}\tau} 
[\alpha_{(2n)1} \nu_1(\tau)+ ... +\alpha_{(2n)n} \nu_n(\tau)] \right).
\label{Eulb}
\ee
These are Gaussian random numbers whose correlations are easily evaluated 
from those of the $\nu_i$'s, 
\be
A_{ij}(t)=<x_i(t)x_j(t)>=\int_0^t d\tau \, e^{-(a_i+a_j)\tau} \, 
(\alpha_{i1}\alpha_{j1} + ... +\alpha_{in} \alpha_{jn})=
\,\frac{1-e^{-t(a_i+a_j)}}{a_i+a_j}\sum_{k=1}^n\alpha_{ik}\alpha_{jk}.
\label{corra}
\ee    
In a more compact matrix notation,
\be
A(t)=<\left[\matrix{x_1(t) \cr \dot{.} \cr x_{2n}(t)}\right] \cdot 
[x_1(t),..,x_{2n}(t)]> = \int_0^t d\tau\, e^{-\tau D}\, \alpha \alpha^T \, e^{-\tau D}.
\label{A}
\ee
Notice the occurrence of the matrix 
$\alpha\alpha^T=\theta^{-1}\left[\matrix{0 & 0 \cr 0 &
2T\beta}\right](\theta^T)^{-1},$
which is trivially positive definite if $\theta$ is real.

Using functional integral techniques, the full distribution function can 
then be easily evaluated (see Appendix B for details) and reads,
\be
W(x_1,...,x_{2n},t;X_{10},...,X_{(2n)0}) = \frac1{(2\pi)^n} \,
\frac1{\sqrt{\det A(t)}} \,
\exp\left(-\frac12 \,[x_1,..,x_{2n}]\, A^{-1}(t) \left[\matrix{x_1 \cr \dot{.} 
\cr x_{2n}}\right] \right),
\ee
where the matrix elements of $A(t)$ are those defined by Eq.(\ref{corra}). 
This makes also a Gaussian distribution, naturally. 

If necessary, it is then 
easy to return to the original variables, $(q_1,...,q_n ; p_1, ... , p_n)$.
Since
\be
\left[\matrix{q_1 \cr \dot{.} \cr q_n \cr p_1 \cr \dot{.} \cr p_n}\right] 
=\theta \left[\matrix{x_1\, e^{a_1t}\cr \dot{.} \cr \dot{.} \cr \dot{.} \cr 
\dot{.}\cr x_{2n}\, e^{a_{2n} t} }\right] + 
\theta \left[\matrix{X_{10}\, e^{a_1t} \cr \dot{.} \cr \dot{.}\cr \dot{.} \cr 
\dot{.} \cr X_{2n0}\, e^{a_{2n}t}} \right], 
\ee 
where the second term of the r.h.s. is the average trajectory and the first 
one is the diffusion part, it is obvious that,
\be
\left[\matrix{<q_1(t)> \cr \dot{.} \cr <q_n(t)> \cr <p_1(t)> \cr
\dot{.} \cr <p_n(t)>}\right] = e^{t{\cal D}}
\left[\matrix{q_{10} \cr \dot{.} \cr q_{n0} \cr p_{10} \cr \dot{.} \cr
p_{n0}}\right] ,
\label{av}\ee
and
\beqn
{\cal A}(t) &=& \left<\left[\matrix{q_1-<q_1(t)> \cr \dot{.} \cr
 q_n-<q_n(t)> \cr
p_1-<p_1(t)> \cr \dot{.} \cr p_n-<p_n(t)>}\right]\cdot [q_1-<q_1(t)>,
..,q_n-<q_n(t)>, p_1-<p_1(t)>,.., p_n-<p_n(t)>]\right>\\
 &=&\theta e^{tD} A(t) e^{tD} \theta^T 
=2T\int_0^t d\tau\, e^{(t-\tau){\cal D}}\,
\left[\matrix{0 & 0 \cr 0 &\beta}\right]\,
e^{(t-\tau){\cal D}^T}.
\label{diff}\eeqn
Eventually, the full distribution function reads,
\be
W(q_1,..,p_n,t;q_{10},..,p_{n0}) = \frac1{(2\pi)^n} \,
\frac1{\sqrt{\det {\cal A}(t)}} \,
\exp\left(-\frac12 [q_1-<q_1(t)>,.., p_n-<p_n(t)>]{\cal A}^{-1}(t)
\left[\matrix{q_1-<q_1(t)> \cr \dot{.} \cr p_n-<p_n(t)>}\right]\right),
\ee
after renormalisation with the Jacobian. This result is well known,
see e.g. \cite{Wang}.

\subsection{Probability to pass over the saddle}

To evaluate the probability of passing over the barrier, we are interested 
in the ``reduced'' distribution obtained when all degrees of freedom but 
$q_1$ are integrated out. It is also necessarily a Gaussian distribution,
\be
W(q_1,t;q_{10},...,q_{n0},p_{10},...,p_{n0})=\frac1{\sqrt{2\pi}\,
\sigma_{q_1}(t)} \exp
\left[-\frac{(q_1-<q_1(t)>)^2}{2\,\sigma_{q_1}^2(t)}
\right]. 
\ee
The only remaining task is then to evaluate $<q_1(t)>$ and
$\sigma_{q_1}(t)$. From Eqs (\ref{av},\ref{diff}) one gets :
\be
<q_1(t)>= \theta_{11}\, X_{10}\, e^{a_1t} + ... + 
\theta_{1(2n)}\, X_{(2n)0}\, e^{a_{2n}t},
\label{q1}
\ee
and
\beqn
\sigma^2_{q_1}(t)&=& <(\theta_{11}\, x_1\, e^{a_1t} + ... + 
\theta_{1(2n)}\, x_{2n}\, e^{a_{2n}t})^2>   \\
 &=& \sum_{i=1}^{2n} \sum_{j=1}^{2n} \theta_{1i}\, \theta_{1j}\, A_{ij}(t)\,
 e^{(a_i+a_j)t} = 
\sum_{i=1}^{2n} \sum_{j=1}^{2n} \theta_{1i} \, 
\frac{e^{(a_i+a_j)t}-1}{a_i+a_j} \, \theta_{1j} 
 \, (\alpha\alpha^T)_{ij} \, \label{sigma1} \\
 &=& 2T\sum_{i=1}^{2n} \sum_{j=1}^{2n} \theta_{1i} \, 
\frac{e^{(a_i+a_j)t}-1}{a_i+a_j} \, \theta_{1j} 
\sum_{v=1}^n \sum_{w=1}^n (\theta^{-1})_{i,n+v} \, \beta_{vw}
 (\theta^{-1})_{i,n+w}.
\label{sigmb1}
\eeqn

To go further and do physics, one needs the eigenvalues and the eigenvectors 
of the DM. This is not always feasible analytically for any dimension. 
However, the previous scheme can be applied to particular problems where the 
drift matrix is explicitly known.

Let us first consider the 1-D and the 2-D cases, for which some general 
features will be shown.

%%%%%%%%%%%%%%%%%%% The 1-D diffusion model

\section{The one dimension problem}
\label{1d}

The 1-D approach is interesting because of its simplicity. Intuitively,
choosing as unique variable, the valley one and averaging all over the others 
should be enough in a first approximation. In this approach, the DM can be 
easily diagonalized and the diffusion energetical condition, time and 
probability can be easily calculated.

\subsection{Solution of the Langevin equation}
There are only one mass and one random force, $m_1$ and $R_1,$ respectively. 
The latter is related to one random number $\nu_1,$ normalized according to 
Eq.(\ref{norma}). The corresponding matrix $\Gamma$ boils down to one 
number only, which, according to the fluctuation-dissipation theorem, reads 
$\Gamma=(2\beta_1 T/m_1)^{-1/2}.$ The eigenvalues of the DM are 
$a=(\beta'_1-\beta_1)/2$ and $b=-(\beta'_1+\beta_1)/2$, with
$\beta'_1=(\beta_1^2+4\omega_1^2)^{1/2}$. Note that $a>0$ and $b<0$.
The matrix $\theta$ and its inverse read,
\be
\theta=(\beta'_1)^{-1}\left[\matrix
{ 1 & -1 \cr (\beta'_1-\beta_1)/2  & (\beta'_1+ \beta_1)/2  }
\right], \ \ \ \  \theta^{-1}=\left[\matrix
{ (\beta'_1+\beta_1)/2 & 1 \cr (\beta_1-\beta'_1)/2 & 1 }
\right]
\, .
\label{thet}
\ee
The matrix $\alpha=\theta^{-1}\left[\matrix{0 \cr \Gamma}\right]$ is thus, 
\be
\alpha=(2T\beta_1/m_1)^{1/2}\left[\matrix{ 1 \cr 1 }\right].
\ee
The eigencoordinates read, with 
$p_1 = \dot q_1,$
\be
\left[\matrix{X \cr Y}\right] = 
\left[ \matrix{ (\beta'_1+\beta_1)/2 & 1 \cr (\beta_1-\beta'_1)/2 & 1} \right] 
\left[\matrix{q_1 \cr p_1}\right]
\, .
\label{thetm1}
\ee
The Euler type variables, $x$ and $y$, are then defined as,
\be
\left\{\begin{array}{ll} x(t) = X e^{-at}-X_0\\ y(t) = Y e^{-bt}-Y_0 
\end{array}\right.,
\ee
and their statistical properties must be the same as those of 
\be
(2T\beta_1/m_1)^{1/2}\left(\int_0^t d\tau \, e^{-a\tau} \nu_1(\tau), 
\int_0^t d\tau \, e^{-b\tau}\nu_1(\tau)\right),
\ee 
see Eqs.(\ref{Eula}-\ref{Eulb}). The random number $\nu_1(t)$ being 
Gaussian, $x$ and $y$ are also Gaussian random variables, with
\be 
<x^2(t)>= \frac{T\beta_1}{a m_1}(1-e^{-2at}), \ \ \ \ 
<y^2(t)>= \frac{T\beta_1}{b m_1}(1-e^{-2bt}), \ \ \ \ 
<x(t)y(t)>= \frac{2T\beta_1}{(a+b)m_1}\left[1-e^{-(a+b)t}\right]
\, ,
\label{corrb}
\ee
see Eq.(\ref{corra}).

\medskip
To evaluate the probability for passing over the barrier, one needs the 
following distribution function, necessarily a Gaussian in the present model,
\be
W(q_1,t;q_{10},p_{10})= \frac1{\sqrt{2\pi} \, \sigma_{q_1}(t)} 
\exp -\frac{[q_1-<q_1(t)>]^2}{2 \, \sigma_{q_1}^2(t)}. 
\label{dist}
\ee
According to the first row of $\theta,$ see Eq.(\ref{thet}), the
valley coordinate is, in terms of the eigencoordinates,
\be
q_1(t)=\frac1{\beta'_1}(x e^{at}- y e^{bt}) + \frac1{\beta'_1}(X_0e^{at} - 
Y_0 e^{bt}).
\ee
The first part corresponds to the diffusion and the second one to the 
average trajectory. It is trivial to obtain $X_0$ and $Y_0$ from
$q_{10}$ and $p_{10}$ according to Eq.(\ref{thetm1}), hence $<q_1(t)>.$ It is 
also trivial to obtain $<q_1^2(t)>$ from Eq.(\ref{corrb}). All told, elementary
manipulations yield,
\be
<q_1(t)>= q_{10} e^{-\beta_1 t/2} 
\left[\cosh\left(\frac12 \beta'_1 t\right) + \frac{\beta_1}{\beta'_1}
\sinh\left(\frac12 \beta'_1 t\right) \right] + 2\,
\frac{p_{10}}{\beta'_1} e^{-\beta_1 t/2} 
\sinh\left(\frac12 \beta'_1 t\right),
\label{xav}
\ee
and 
\be
\sigma_{q_1}^2(t)= <q_1^2(t)>-<q_1(t)>^2= -\frac{T}{m_1\omega_1^2}
\left[1-e^{-\beta_1 t}\left(2 \frac{\beta_1^2}{\beta_1^{'2}} 
\sinh^2\left(\frac12\beta'_1 t\right) + \frac{\beta_1}{\beta'_1}
\sinh\left(\beta'_1 t\right) + 1\right)\right].
\label{sigma}
\ee
These results are in agreement with the well-known calculation done for 
harmonically bound particles \cite{Uhl,Chan}, where the sign of the spring 
force should be changed.

\subsection{The critical initial kinetic energy}

Defining the critical energy as the kinetic energy, 
$K=\frac12 m_1 p_{10}^2$, necessary to have half of the particles to
pass over the barrier, it is obvious that it corresponds to
$\lim_{t\rightarrow +\infty}<q_1(t)>=0$.
From Eq.(\ref{xav}) it can be easily shown that 
\be
K_{c} = \left(\frac{\beta_1+\beta'_1}{2\omega_1}\right)^2 B,
\label{extra}
\ee
where $B=m_1 \omega_1^2q_{10}^2/2$ is the barrier height. 
In the weak friction limit, $\beta_1 \simeq 0$, this condition 
becomes $K_{c}=B$ which is a trivial result.  In the overdamped limit, 
$\beta_1 \gg 2\omega_1$, it becomes, $K_{c} = (\beta_1/\omega_1)^2 B$.

In the case of nuclei, using typical values, $\hbar\omega_1=1MeV$ and 
$\beta_1=5. 10^{21}s^{-1}$, the overdamped limit is usually a good 
approximation and a big kinetic energy is necessary to overpass even 
a very small barrier: $K_c\simeq10 B$.

%%%%%%%%%%%%%%%%%%%%%%%%%%%

\subsection{Diffusion time}

As for many physical problems the diffusion process dynamically competes 
with some other processes, it is interesting to extract also the time 
necessary to reach the top of the potential barrier. When the previous 
condition, Eq.(\ref{extra}), is exactly fulfilled, then
\be
<q_1(t)>= q_{10}\, e^{-\frac{\beta_1+\beta_1'}2 t}.
\ee
The average trajectory exponentially converges to the top of the barrier 
with a typical time equal to $2/(\beta_1+\beta_1')$, which becomes 
$1/\omega_1$ in the weak damping limit and $1/\beta_1$ in the overdamped one.

When the initial kinetic energy is higher than the critical one,
Eq.(\ref{extra}), the average trajectory reaches the top of the potential
barrier at $t_{top}$ such as,
\be
\coth\left(\frac12\, \beta_1'\, t_{top}\right) = \frac{2\omega_1}{\beta_1'}
\left(\sqrt{\frac{K}{B}}-\frac{\beta_1}{2\omega_1}\right).
\ee
The previous equation becomes
\beqn
\coth(\omega_1\, t_{top}) &=& \sqrt{\frac{K}{B}} \qquad {\rm in\ the\ weak
\ damping\ case,}\\
\coth\left(\frac12\, \beta_1\, t_{top}\right) &=& \left(\frac{2\omega_1}
{\beta_1} \sqrt{\frac{K}{B}} - 1\right) \quad {\rm in\ the\ overdamped\ one.}
\eeqn

%%%%%%%%%%%%%%%%%%%%%%%%%%%%%%%%%

\subsection{Passing probability}

In this model the probability at a given time that the particle has passed 
over the barrier is simply,
\beqn
P(t;q_{10},p_{10}) &=& \int_0^{+\infty} W(q_1,t;q_{10},p_{10})\, dq_1\\
             &=& \frac12\, {\rm erfc} 
\left(-\frac{<q_1(t)>}{\sqrt2 \sigma_{q_1}(t)}\right).
\label{passed}\eeqn

For large times ($t\gg1/\beta_1'$),
\beqn
-\frac{<q_1>}{\sqrt2 \sigma_{q_1}} &\rightarrow & \frac{\beta_1+\beta_1'}
{\sqrt{2(\beta_1^2+\beta_1\beta_1')}} \left[\sqrt{\frac{B}{T}}- 
\frac{2\omega_1}{\beta_1+\beta_1'} \sqrt{\frac{K}{T}} \right] 
\label{lt}\\
 &\rightarrow & \sqrt{\frac{\omega_1}{T\beta_1}} (\sqrt{B} - \sqrt{K}) 
\quad {\rm in\ the\ underdamped\ limit,}\\
 &\rightarrow & \sqrt{\frac{B}{T}} - \frac{\omega_1}{\beta_1} 
\sqrt{\frac{K}{T}} \quad {\rm in\ the\ overdamped\ one}.
\eeqn
The passing probability is then known as a function of the initial kinetic 
energy and the temperature. It increases from 0 to 1 around the critical 
value $K_c$ when increasing the initial kinetic energy. The higher the 
temperature, the smoother is this increase.

%%%%%%%%%%%%%%%%%%%%%%%%%%%%%%%%%

\subsection{Heavy-ion fusion}

When two nuclei are colliding, the kinetic energy of the projectile should 
be higher than the contact energy derived by Bass \cite{Bass} to observe 
some fusion events. This so-called extra-push energy is generally 
interpreted as an additional barrier due to nuclear forces that has to be
overcome by the viscous nuclear matter \cite{Bjo}. The critical energy 
derived above can then be seen as the extra-push energy here.

To calculate the fusion probability of two cold colliding nuclei, the 
difficulty is then to evaluate the temperature. For particles that 
can reach the barrier top, at a distance $R_{12},$ we will assume that all 
the remaining energy is totally dissipated. Therefore, with a level 
density $a_{lev},$ we assume that $a_{lev} T^2 = K-B$, neglecting the 
collective kinetic energy. Then the fusion probability is,
\be
P(q_{10},p_{10}) = \frac12\, {\rm erfc} 
\left[\left(\frac{a_{lev}B^2}{K-B}\right)^{1/4}
\frac{\beta_1+\beta_1'}{\sqrt{2(\beta_1^2+\beta_1\beta_1')}}
\left(1-\frac{2\omega_1}{\beta_1+\beta_1'}\sqrt{\frac{K}{B}}\right)\right].
\ee
In this formula, $K$ is the available kinetic energy when the two nuclei 
are in contact, and $B$ is the remaining barrier height:
\be
K= E_{cm} - E_{rot} - B_{Bass}.
\ee
In this equation, $B_{Bass}$ is the Bass barrier \cite{Bass} and $E_{rot}$ 
is the rotational energy, $E_{rot} = E_{cm} b_{imp}^2/R_{12}^2$, 
where $b_{imp}$ is the impact parameter. 

Therefore, the fusion probability reads,
\be
P(E_{cm},b)=\frac12\, {\rm erfc}\left[\left(\frac{a_{lev}B^2}{E_{cm}
(1-\frac{b_{imp}^2}{R_{12}^2})-B_{Bass}-B}\right)^{1/4} 
\frac{\beta_1+\beta_1'}{\sqrt{2(\beta_1^2+\beta_1\beta_1')}}
\left(1-\frac{2\omega_1}{\beta_1+\beta_1'}
\sqrt{\frac{E_{cm}(1-\frac{b_{imp}^2}{R_{12}^2})-B_{Bass}}{B}}\right)\right].
\ee

To compare this probability with other theoretical calculations, we need to
choose parameters specific to our problem and deduce the potential and
dissipation terms. The present paper being rather dedicated to a good 
qualitative understanding of the Langevin model, numerics with realistic 
parameters will be kept for a future paper. However, numerical 
simulations already available seem to indicate that at least two dimensions
are necessary to really have a good understanding of the diffusion phenomena 
in this problem \cite{Agu}.

%%%%%%%%%%%%%%%%%%%%%%%%%%%% The 2-D diffusion model

\section{The two-dimensional model}

% Full 2-D equation

\subsection{Passing probability}

Choosing $q_1$ as the valley variable, the associated distribution function, 
$W(q_1,t;q_{10},q_{20},p_{10},p_{20})$, is a Gaussian and the probability 
at a given time to pass over the saddle reads,
\beqn
P(t;q_{10},q_{20},p_{10},p_{20}) &=& \int_0^{+\infty} 
W(q_1,t;q_{10},q_{20},p_{10},p_{20})\, dq_1\\
             &=& \frac12\, {\rm erfc} 
\left(-\frac{<q_1(t)>}{\sqrt2 \, \sigma_{q_1}(t)}\right).
\eeqn
The difficulty is then to find $<q_1(t)>$ and $\sigma_{q_1}(t)$, see 
Eqs.(\ref{q1}) and (\ref{sigma1}). 
When the two degrees of freedom are uncorrelated, namely when $\beta_{12}=0,$
the eigenvalues of the DM are, obviously,
\be
\left\{\begin{array}{llll} a_1=(-\beta_1+\sqrt{\beta_1^2+4\omega_1^2})/2\\
a_2=(-\beta_1-\sqrt{\beta_1^2+4\omega_1^2})/2\\
a_3=(-\beta_2+\sqrt{\beta_2^2-4\omega_2^2})/2\\
a_4=(-\beta_2-\sqrt{\beta_2^2-4\omega_2^2})/2\\ 
\end{array}\right.
\quad  {\rm or} \quad
\left\{\begin{array}{llll} a_1=(-\beta_1+\sqrt{\beta_1^2+4\omega_1^2})/2\\
a_2=(-\beta_1-\sqrt{\beta_1^2+4\omega_1^2})/2\\
a_3=(-\beta_2+i\sqrt{-\beta_2^2+4\omega_2^2})/2\\
a_4=(-\beta_2-i\sqrt{-\beta_2^2+4\omega_2^2})/2\\ 
\end{array}\right.,
\label{eigen}
\ee
if $\omega_2<\beta_2/2$ or $\omega_2>\beta_2/2,$ respectively. Note that 
only $a_1$ is positive, the other eigenvalues are negative or have a 
negative real part.

When $\beta_{12} \ne 0,$ it is easy to show that there are always two 
real roots. The first one, $a_1,$ is positive and increasing as a function 
of $\beta_{12}.$ The other one, $a_4,$ is negative and decreasing as a 
function of the same. When the other two roots are real, it is also 
trivial to show that they remain negative. For large values 
of $\beta_{12},$ which start reaching significant fractions of that 
maximum, $(\beta_1 \beta_2)^{1/2},$ which is acceptable for a semipositivity 
of friction, the other two roots may become complex conjugate. But the
statement $\Re(a_i)<0$ for $i>1,$ is still true. To illustrate the
behavior of the eigenvalues, we consider the special case where
$\beta_1=\beta_2=1$ and $\omega_1=\omega_2=\omega.$ The equal diagonal 
viscosities being taken as a unit, the polynomial equation whose roots are 
the eigenfrequencies of the problem reads,
\be
a^4+2a^3+a^2-\omega^4=\beta_{12}^2a^2.
\ee

Fig.1 shows, when $\beta_{12}$ increases from $0$ to $1,$ the graphs
of the four roots, or of their real parts when some of them become
complex. Here $\omega=0.35,$ which corresponds to overdamping at the 
beginning, when $\beta_{12}=0.$ The merging of the two intermediate 
roots and their complexification are transparent. Once such roots have 
become complex conjugate, their common real part, however, remains negative.

\begin{figure}[h] \centering
%\mbox{  \epsfysize=100mm \epsffile{figboi1.eps} }
% Fig.1 of EH7154 Abe
%%%%%%%%%%%%%%%%
% GNUPLOT: LaTeX picture
\setlength{\unitlength}{0.240900pt}
\ifx\plotpoint\undefined\newsavebox{\plotpoint}\fi
\begin{picture}(1500,900)(0,0)
\font\gnuplot=cmr10 at 10pt
\gnuplot
\sbox{\plotpoint}{\rule[-0.200pt]{0.400pt}{0.400pt}}%
\put(220.0,724.0){\rule[-0.200pt]{292.934pt}{0.400pt}}
\put(220.0,113.0){\rule[-0.200pt]{0.400pt}{184.048pt}}
\put(220.0,113.0){\rule[-0.200pt]{4.818pt}{0.400pt}}
\put(198,113){\makebox(0,0)[r]{-2}}
\put(1416.0,113.0){\rule[-0.200pt]{4.818pt}{0.400pt}}
\put(220.0,266.0){\rule[-0.200pt]{4.818pt}{0.400pt}}
\put(198,266){\makebox(0,0)[r]{-1.5}}
\put(1416.0,266.0){\rule[-0.200pt]{4.818pt}{0.400pt}}
\put(220.0,419.0){\rule[-0.200pt]{4.818pt}{0.400pt}}
\put(198,419){\makebox(0,0)[r]{-1}}
\put(1416.0,419.0){\rule[-0.200pt]{4.818pt}{0.400pt}}
\put(220.0,571.0){\rule[-0.200pt]{4.818pt}{0.400pt}}
\put(198,571){\makebox(0,0)[r]{-0.5}}
\put(1416.0,571.0){\rule[-0.200pt]{4.818pt}{0.400pt}}
\put(220.0,724.0){\rule[-0.200pt]{4.818pt}{0.400pt}}
\put(198,724){\makebox(0,0)[r]{0}}
\put(1416.0,724.0){\rule[-0.200pt]{4.818pt}{0.400pt}}
\put(220.0,877.0){\rule[-0.200pt]{4.818pt}{0.400pt}}
\put(198,877){\makebox(0,0)[r]{0.5}}
\put(1416.0,877.0){\rule[-0.200pt]{4.818pt}{0.400pt}}
\put(220.0,113.0){\rule[-0.200pt]{0.400pt}{4.818pt}}
\put(220,68){\makebox(0,0){0}}
\put(220.0,857.0){\rule[-0.200pt]{0.400pt}{4.818pt}}
\put(530.0,113.0){\rule[-0.200pt]{0.400pt}{4.818pt}}
\put(530,68){\makebox(0,0){0.25}}
\put(530.0,857.0){\rule[-0.200pt]{0.400pt}{4.818pt}}
\put(840.0,113.0){\rule[-0.200pt]{0.400pt}{4.818pt}}
\put(840,68){\makebox(0,0){0.5}}
\put(840.0,857.0){\rule[-0.200pt]{0.400pt}{4.818pt}}
\put(1151.0,113.0){\rule[-0.200pt]{0.400pt}{4.818pt}}
\put(1151,68){\makebox(0,0){0.75}}
\put(1151.0,857.0){\rule[-0.200pt]{0.400pt}{4.818pt}}
\put(220.0,113.0){\rule[-0.200pt]{292.934pt}{0.400pt}}
\put(1436.0,113.0){\rule[-0.200pt]{0.400pt}{184.048pt}}
\put(220.0,877.0){\rule[-0.200pt]{292.934pt}{0.400pt}}
\put(170,820){\makebox(0,0){$Re \, a$}}
\put(1390,75){\makebox(0,0){$\beta_{12}$}}

\put(245,384){\raisebox{-.8pt}{\makebox(0,0){\circle*{12}}}}
\put(245,463){\raisebox{-.8pt}{\makebox(0,0){\circle*{12}}}}
\put(245,680){\raisebox{-.8pt}{\makebox(0,0){\circle*{12}}}}
\put(245,758){\raisebox{-.8pt}{\makebox(0,0){\circle*{12}}}}
\put(270,383){\raisebox{-.8pt}{\makebox(0,0){\circle*{12}}}}
\put(270,464){\raisebox{-.8pt}{\makebox(0,0){\circle*{12}}}}
\put(270,680){\raisebox{-.8pt}{\makebox(0,0){\circle*{12}}}}
\put(270,758){\raisebox{-.8pt}{\makebox(0,0){\circle*{12}}}}
\put(294,381){\raisebox{-.8pt}{\makebox(0,0){\circle*{12}}}}
\put(294,467){\raisebox{-.8pt}{\makebox(0,0){\circle*{12}}}}
\put(294,680){\raisebox{-.8pt}{\makebox(0,0){\circle*{12}}}}
\put(294,758){\raisebox{-.8pt}{\makebox(0,0){\circle*{12}}}}
\put(319,378){\raisebox{-.8pt}{\makebox(0,0){\circle*{12}}}}
\put(319,470){\raisebox{-.8pt}{\makebox(0,0){\circle*{12}}}}
\put(319,680){\raisebox{-.8pt}{\makebox(0,0){\circle*{12}}}}
\put(319,758){\raisebox{-.8pt}{\makebox(0,0){\circle*{12}}}}
\put(344,374){\raisebox{-.8pt}{\makebox(0,0){\circle*{12}}}}
\put(344,474){\raisebox{-.8pt}{\makebox(0,0){\circle*{12}}}}
\put(344,680){\raisebox{-.8pt}{\makebox(0,0){\circle*{12}}}}
\put(344,758){\raisebox{-.8pt}{\makebox(0,0){\circle*{12}}}}
\put(369,370){\raisebox{-.8pt}{\makebox(0,0){\circle*{12}}}}
\put(369,478){\raisebox{-.8pt}{\makebox(0,0){\circle*{12}}}}
\put(369,680){\raisebox{-.8pt}{\makebox(0,0){\circle*{12}}}}
\put(369,758){\raisebox{-.8pt}{\makebox(0,0){\circle*{12}}}}
\put(394,365){\raisebox{-.8pt}{\makebox(0,0){\circle*{12}}}}
\put(394,482){\raisebox{-.8pt}{\makebox(0,0){\circle*{12}}}}
\put(394,680){\raisebox{-.8pt}{\makebox(0,0){\circle*{12}}}}
\put(394,758){\raisebox{-.8pt}{\makebox(0,0){\circle*{12}}}}
\put(419,361){\raisebox{-.8pt}{\makebox(0,0){\circle*{12}}}}
\put(419,487){\raisebox{-.8pt}{\makebox(0,0){\circle*{12}}}}
\put(419,680){\raisebox{-.8pt}{\makebox(0,0){\circle*{12}}}}
\put(419,758){\raisebox{-.8pt}{\makebox(0,0){\circle*{12}}}}
\put(443,355){\raisebox{-.8pt}{\makebox(0,0){\circle*{12}}}}
\put(443,493){\raisebox{-.8pt}{\makebox(0,0){\circle*{12}}}}
\put(443,679){\raisebox{-.8pt}{\makebox(0,0){\circle*{12}}}}
\put(443,758){\raisebox{-.8pt}{\makebox(0,0){\circle*{12}}}}
\put(468,350){\raisebox{-.8pt}{\makebox(0,0){\circle*{12}}}}
\put(468,498){\raisebox{-.8pt}{\makebox(0,0){\circle*{12}}}}
\put(468,679){\raisebox{-.8pt}{\makebox(0,0){\circle*{12}}}}
\put(468,758){\raisebox{-.8pt}{\makebox(0,0){\circle*{12}}}}
\put(493,345){\raisebox{-.8pt}{\makebox(0,0){\circle*{12}}}}
\put(493,504){\raisebox{-.8pt}{\makebox(0,0){\circle*{12}}}}
\put(493,679){\raisebox{-.8pt}{\makebox(0,0){\circle*{12}}}}
\put(493,758){\raisebox{-.8pt}{\makebox(0,0){\circle*{12}}}}
\put(518,339){\raisebox{-.8pt}{\makebox(0,0){\circle*{12}}}}
\put(518,509){\raisebox{-.8pt}{\makebox(0,0){\circle*{12}}}}
\put(518,678){\raisebox{-.8pt}{\makebox(0,0){\circle*{12}}}}
\put(518,759){\raisebox{-.8pt}{\makebox(0,0){\circle*{12}}}}
\put(543,334){\raisebox{-.8pt}{\makebox(0,0){\circle*{12}}}}
\put(543,515){\raisebox{-.8pt}{\makebox(0,0){\circle*{12}}}}
\put(543,678){\raisebox{-.8pt}{\makebox(0,0){\circle*{12}}}}
\put(543,759){\raisebox{-.8pt}{\makebox(0,0){\circle*{12}}}}
\put(567,328){\raisebox{-.8pt}{\makebox(0,0){\circle*{12}}}}
\put(567,521){\raisebox{-.8pt}{\makebox(0,0){\circle*{12}}}}
\put(567,677){\raisebox{-.8pt}{\makebox(0,0){\circle*{12}}}}
\put(567,759){\raisebox{-.8pt}{\makebox(0,0){\circle*{12}}}}
\put(592,323){\raisebox{-.8pt}{\makebox(0,0){\circle*{12}}}}
\put(592,527){\raisebox{-.8pt}{\makebox(0,0){\circle*{12}}}}
\put(592,677){\raisebox{-.8pt}{\makebox(0,0){\circle*{12}}}}
\put(592,759){\raisebox{-.8pt}{\makebox(0,0){\circle*{12}}}}
\put(617,317){\raisebox{-.8pt}{\makebox(0,0){\circle*{12}}}}
\put(617,533){\raisebox{-.8pt}{\makebox(0,0){\circle*{12}}}}
\put(617,676){\raisebox{-.8pt}{\makebox(0,0){\circle*{12}}}}
\put(617,759){\raisebox{-.8pt}{\makebox(0,0){\circle*{12}}}}
\put(642,311){\raisebox{-.8pt}{\makebox(0,0){\circle*{12}}}}
\put(642,540){\raisebox{-.8pt}{\makebox(0,0){\circle*{12}}}}
\put(642,676){\raisebox{-.8pt}{\makebox(0,0){\circle*{12}}}}
\put(642,759){\raisebox{-.8pt}{\makebox(0,0){\circle*{12}}}}
\put(667,305){\raisebox{-.8pt}{\makebox(0,0){\circle*{12}}}}
\put(667,546){\raisebox{-.8pt}{\makebox(0,0){\circle*{12}}}}
\put(667,675){\raisebox{-.8pt}{\makebox(0,0){\circle*{12}}}}
\put(667,760){\raisebox{-.8pt}{\makebox(0,0){\circle*{12}}}}
\put(692,299){\raisebox{-.8pt}{\makebox(0,0){\circle*{12}}}}
\put(692,552){\raisebox{-.8pt}{\makebox(0,0){\circle*{12}}}}
\put(692,674){\raisebox{-.8pt}{\makebox(0,0){\circle*{12}}}}
\put(692,760){\raisebox{-.8pt}{\makebox(0,0){\circle*{12}}}}
\put(716,294){\raisebox{-.8pt}{\makebox(0,0){\circle*{12}}}}
\put(716,559){\raisebox{-.8pt}{\makebox(0,0){\circle*{12}}}}
\put(716,673){\raisebox{-.8pt}{\makebox(0,0){\circle*{12}}}}
\put(716,760){\raisebox{-.8pt}{\makebox(0,0){\circle*{12}}}}
\put(741,287){\raisebox{-.8pt}{\makebox(0,0){\circle*{12}}}}
\put(741,566){\raisebox{-.8pt}{\makebox(0,0){\circle*{12}}}}
\put(741,672){\raisebox{-.8pt}{\makebox(0,0){\circle*{12}}}}
\put(741,760){\raisebox{-.8pt}{\makebox(0,0){\circle*{12}}}}
\put(766,282){\raisebox{-.8pt}{\makebox(0,0){\circle*{12}}}}
\put(766,573){\raisebox{-.8pt}{\makebox(0,0){\circle*{12}}}}
\put(766,670){\raisebox{-.8pt}{\makebox(0,0){\circle*{12}}}}
\put(766,761){\raisebox{-.8pt}{\makebox(0,0){\circle*{12}}}}
\put(791,276){\raisebox{-.8pt}{\makebox(0,0){\circle*{12}}}}
\put(791,580){\raisebox{-.8pt}{\makebox(0,0){\circle*{12}}}}
\put(791,669){\raisebox{-.8pt}{\makebox(0,0){\circle*{12}}}}
\put(791,761){\raisebox{-.8pt}{\makebox(0,0){\circle*{12}}}}
\put(816,270){\raisebox{-.8pt}{\makebox(0,0){\circle*{12}}}}
\put(816,588){\raisebox{-.8pt}{\makebox(0,0){\circle*{12}}}}
\put(816,667){\raisebox{-.8pt}{\makebox(0,0){\circle*{12}}}}
\put(816,761){\raisebox{-.8pt}{\makebox(0,0){\circle*{12}}}}
\put(840,264){\raisebox{-.8pt}{\makebox(0,0){\circle*{12}}}}
\put(840,595){\raisebox{-.8pt}{\makebox(0,0){\circle*{12}}}}
\put(840,665){\raisebox{-.8pt}{\makebox(0,0){\circle*{12}}}}
\put(840,761){\raisebox{-.8pt}{\makebox(0,0){\circle*{12}}}}
\put(865,258){\raisebox{-.8pt}{\makebox(0,0){\circle*{12}}}}
\put(865,604){\raisebox{-.8pt}{\makebox(0,0){\circle*{12}}}}
\put(865,662){\raisebox{-.8pt}{\makebox(0,0){\circle*{12}}}}
\put(865,762){\raisebox{-.8pt}{\makebox(0,0){\circle*{12}}}}
\put(890,252){\raisebox{-.8pt}{\makebox(0,0){\circle*{12}}}}
\put(890,613){\raisebox{-.8pt}{\makebox(0,0){\circle*{12}}}}
\put(890,658){\raisebox{-.8pt}{\makebox(0,0){\circle*{12}}}}
\put(890,762){\raisebox{-.8pt}{\makebox(0,0){\circle*{12}}}}
\put(915,246){\raisebox{-.8pt}{\makebox(0,0){\circle*{12}}}}
\put(915,625){\raisebox{-.8pt}{\makebox(0,0){\circle*{12}}}}
\put(915,652){\raisebox{-.8pt}{\makebox(0,0){\circle*{12}}}}
\put(915,762){\raisebox{-.8pt}{\makebox(0,0){\circle*{12}}}}
\put(940,240){\raisebox{-.8pt}{\makebox(0,0){\circle*{12}}}}
\put(940,641){\raisebox{-.8pt}{\makebox(0,0){\circle*{12}}}}
\put(940,641){\raisebox{-.8pt}{\makebox(0,0){\circle*{12}}}}
\put(940,763){\raisebox{-.8pt}{\makebox(0,0){\circle*{12}}}}
\put(964,234){\raisebox{-.8pt}{\makebox(0,0){\circle*{12}}}}
\put(964,644){\raisebox{-.8pt}{\makebox(0,0){\circle*{12}}}}
\put(964,644){\raisebox{-.8pt}{\makebox(0,0){\circle*{12}}}}
\put(964,763){\raisebox{-.8pt}{\makebox(0,0){\circle*{12}}}}
\put(989,228){\raisebox{-.8pt}{\makebox(0,0){\circle*{12}}}}
\put(989,647){\raisebox{-.8pt}{\makebox(0,0){\circle*{12}}}}
\put(989,647){\raisebox{-.8pt}{\makebox(0,0){\circle*{12}}}}
\put(989,764){\raisebox{-.8pt}{\makebox(0,0){\circle*{12}}}}
\put(1014,222){\raisebox{-.8pt}{\makebox(0,0){\circle*{12}}}}
\put(1014,650){\raisebox{-.8pt}{\makebox(0,0){\circle*{12}}}}
\put(1014,650){\raisebox{-.8pt}{\makebox(0,0){\circle*{12}}}}
\put(1014,764){\raisebox{-.8pt}{\makebox(0,0){\circle*{12}}}}
\put(1039,216){\raisebox{-.8pt}{\makebox(0,0){\circle*{12}}}}
\put(1039,653){\raisebox{-.8pt}{\makebox(0,0){\circle*{12}}}}
\put(1039,653){\raisebox{-.8pt}{\makebox(0,0){\circle*{12}}}}
\put(1039,765){\raisebox{-.8pt}{\makebox(0,0){\circle*{12}}}}
\put(1064,210){\raisebox{-.8pt}{\makebox(0,0){\circle*{12}}}}
\put(1064,655){\raisebox{-.8pt}{\makebox(0,0){\circle*{12}}}}
\put(1064,655){\raisebox{-.8pt}{\makebox(0,0){\circle*{12}}}}
\put(1064,765){\raisebox{-.8pt}{\makebox(0,0){\circle*{12}}}}
\put(1089,203){\raisebox{-.8pt}{\makebox(0,0){\circle*{12}}}}
\put(1089,658){\raisebox{-.8pt}{\makebox(0,0){\circle*{12}}}}
\put(1089,658){\raisebox{-.8pt}{\makebox(0,0){\circle*{12}}}}
\put(1089,766){\raisebox{-.8pt}{\makebox(0,0){\circle*{12}}}}
\put(1113,198){\raisebox{-.8pt}{\makebox(0,0){\circle*{12}}}}
\put(1113,661){\raisebox{-.8pt}{\makebox(0,0){\circle*{12}}}}
\put(1113,661){\raisebox{-.8pt}{\makebox(0,0){\circle*{12}}}}
\put(1113,767){\raisebox{-.8pt}{\makebox(0,0){\circle*{12}}}}
\put(1138,192){\raisebox{-.8pt}{\makebox(0,0){\circle*{12}}}}
\put(1138,663){\raisebox{-.8pt}{\makebox(0,0){\circle*{12}}}}
\put(1138,663){\raisebox{-.8pt}{\makebox(0,0){\circle*{12}}}}
\put(1138,767){\raisebox{-.8pt}{\makebox(0,0){\circle*{12}}}}
\put(1163,185){\raisebox{-.8pt}{\makebox(0,0){\circle*{12}}}}
\put(1163,666){\raisebox{-.8pt}{\makebox(0,0){\circle*{12}}}}
\put(1163,666){\raisebox{-.8pt}{\makebox(0,0){\circle*{12}}}}
\put(1163,768){\raisebox{-.8pt}{\makebox(0,0){\circle*{12}}}}
\put(1188,179){\raisebox{-.8pt}{\makebox(0,0){\circle*{12}}}}
\put(1188,669){\raisebox{-.8pt}{\makebox(0,0){\circle*{12}}}}
\put(1188,669){\raisebox{-.8pt}{\makebox(0,0){\circle*{12}}}}
\put(1188,769){\raisebox{-.8pt}{\makebox(0,0){\circle*{12}}}}
\put(1213,173){\raisebox{-.8pt}{\makebox(0,0){\circle*{12}}}}
\put(1213,671){\raisebox{-.8pt}{\makebox(0,0){\circle*{12}}}}
\put(1213,671){\raisebox{-.8pt}{\makebox(0,0){\circle*{12}}}}
\put(1213,770){\raisebox{-.8pt}{\makebox(0,0){\circle*{12}}}}
\put(1237,167){\raisebox{-.8pt}{\makebox(0,0){\circle*{12}}}}
\put(1237,674){\raisebox{-.8pt}{\makebox(0,0){\circle*{12}}}}
\put(1237,674){\raisebox{-.8pt}{\makebox(0,0){\circle*{12}}}}
\put(1237,771){\raisebox{-.8pt}{\makebox(0,0){\circle*{12}}}}
\put(1262,161){\raisebox{-.8pt}{\makebox(0,0){\circle*{12}}}}
\put(1262,677){\raisebox{-.8pt}{\makebox(0,0){\circle*{12}}}}
\put(1262,677){\raisebox{-.8pt}{\makebox(0,0){\circle*{12}}}}
\put(1262,772){\raisebox{-.8pt}{\makebox(0,0){\circle*{12}}}}
\put(1287,155){\raisebox{-.8pt}{\makebox(0,0){\circle*{12}}}}
\put(1287,679){\raisebox{-.8pt}{\makebox(0,0){\circle*{12}}}}
\put(1287,679){\raisebox{-.8pt}{\makebox(0,0){\circle*{12}}}}
\put(1287,772){\raisebox{-.8pt}{\makebox(0,0){\circle*{12}}}}
\put(1312,149){\raisebox{-.8pt}{\makebox(0,0){\circle*{12}}}}
\put(1312,681){\raisebox{-.8pt}{\makebox(0,0){\circle*{12}}}}
\put(1312,681){\raisebox{-.8pt}{\makebox(0,0){\circle*{12}}}}
\put(1312,774){\raisebox{-.8pt}{\makebox(0,0){\circle*{12}}}}
\put(1337,143){\raisebox{-.8pt}{\makebox(0,0){\circle*{12}}}}
\put(1337,684){\raisebox{-.8pt}{\makebox(0,0){\circle*{12}}}}
\put(1337,684){\raisebox{-.8pt}{\makebox(0,0){\circle*{12}}}}
\put(1337,775){\raisebox{-.8pt}{\makebox(0,0){\circle*{12}}}}
\put(1362,137){\raisebox{-.8pt}{\makebox(0,0){\circle*{12}}}}
\put(1362,686){\raisebox{-.8pt}{\makebox(0,0){\circle*{12}}}}
\put(1362,686){\raisebox{-.8pt}{\makebox(0,0){\circle*{12}}}}
\put(1362,776){\raisebox{-.8pt}{\makebox(0,0){\circle*{12}}}}
\put(1386,131){\raisebox{-.8pt}{\makebox(0,0){\circle*{12}}}}
\put(1386,689){\raisebox{-.8pt}{\makebox(0,0){\circle*{12}}}}
\put(1386,689){\raisebox{-.8pt}{\makebox(0,0){\circle*{12}}}}
\put(1386,777){\raisebox{-.8pt}{\makebox(0,0){\circle*{12}}}}
\put(1411,125){\raisebox{-.8pt}{\makebox(0,0){\circle*{12}}}}
\put(1411,691){\raisebox{-.8pt}{\makebox(0,0){\circle*{12}}}}
\put(1411,691){\raisebox{-.8pt}{\makebox(0,0){\circle*{12}}}}
\put(1411,779){\raisebox{-.8pt}{\makebox(0,0){\circle*{12}}}}
\put(1436,119){\raisebox{-.8pt}{\makebox(0,0){\circle*{12}}}}
\put(1436,693){\raisebox{-.8pt}{\makebox(0,0){\circle*{12}}}}
\put(1436,693){\raisebox{-.8pt}{\makebox(0,0){\circle*{12}}}}
\put(1436,780){\raisebox{-.8pt}{\makebox(0,0){\circle*{12}}}}
\end{picture}
%%%%%%%%%%%%%%
\caption{Real part of eigenvalues as functions of non diagonal friction. 
Overdamped regime, $\omega=0.35<\beta/2=0.5.$ Notice how two real roots fuse, 
then become complex conjugate, to generate one branch instead of two.} 
\end{figure}
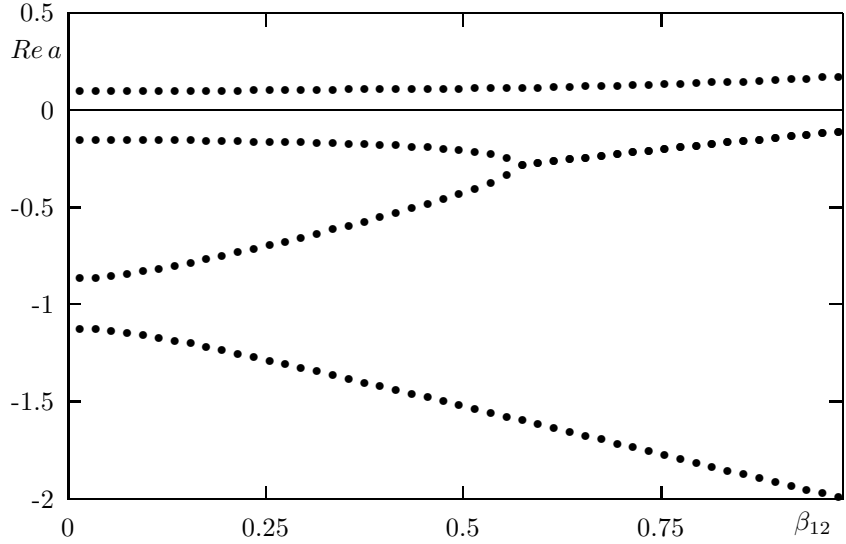

Then Fig.2 shows the case $\omega=0.6,$ where the two intermediate
roots are always complex, with a negative, common real part.

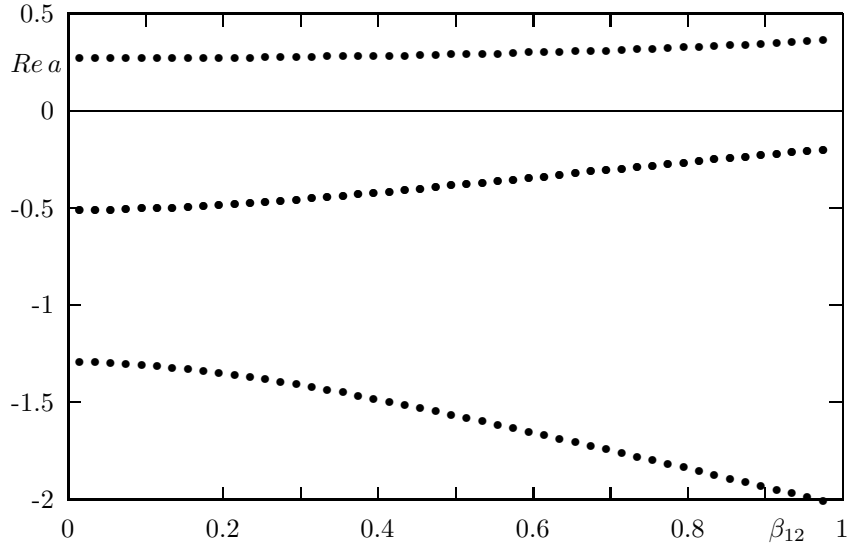
\begin{figure}[h] \centering
%\mbox{  \epsfysize=100mm   \epsffile{figboi2.eps} }
% Fig.2 of EH7154 Abe
%%%%%%%%%%%%%%%
% GNUPLOT: LaTeX picture
\setlength{\unitlength}{0.240900pt}
\ifx\plotpoint\undefined\newsavebox{\plotpoint}\fi
\sbox{\plotpoint}{\rule[-0.200pt]{0.400pt}{0.400pt}}%
\begin{picture}(1500,900)(0,0)
\font\gnuplot=cmr10 at 10pt
\gnuplot
\sbox{\plotpoint}{\rule[-0.200pt]{0.400pt}{0.400pt}}%
\put(220.0,724.0){\rule[-0.200pt]{292.934pt}{0.400pt}}
\put(220.0,113.0){\rule[-0.200pt]{0.400pt}{184.048pt}}
\put(220.0,113.0){\rule[-0.200pt]{4.818pt}{0.400pt}}
\put(198,113){\makebox(0,0)[r]{-2}}
\put(1416.0,113.0){\rule[-0.200pt]{4.818pt}{0.400pt}}
\put(220.0,266.0){\rule[-0.200pt]{4.818pt}{0.400pt}}
\put(198,266){\makebox(0,0)[r]{-1.5}}
\put(1416.0,266.0){\rule[-0.200pt]{4.818pt}{0.400pt}}
\put(220.0,419.0){\rule[-0.200pt]{4.818pt}{0.400pt}}
\put(198,419){\makebox(0,0)[r]{-1}}
\put(1416.0,419.0){\rule[-0.200pt]{4.818pt}{0.400pt}}
\put(220.0,571.0){\rule[-0.200pt]{4.818pt}{0.400pt}}
\put(198,571){\makebox(0,0)[r]{-0.5}}
\put(1416.0,571.0){\rule[-0.200pt]{4.818pt}{0.400pt}}
\put(220.0,724.0){\rule[-0.200pt]{4.818pt}{0.400pt}}
\put(198,724){\makebox(0,0)[r]{0}}
\put(1416.0,724.0){\rule[-0.200pt]{4.818pt}{0.400pt}}
\put(220.0,877.0){\rule[-0.200pt]{4.818pt}{0.400pt}}
\put(198,877){\makebox(0,0)[r]{0.5}}
\put(1416.0,877.0){\rule[-0.200pt]{4.818pt}{0.400pt}}
\put(220.0,113.0){\rule[-0.200pt]{0.400pt}{4.818pt}}
\put(220,68){\makebox(0,0){0}}
\put(342.0,857.0){\rule[-0.200pt]{0.400pt}{4.818pt}}
\put(463.0,113.0){\rule[-0.200pt]{0.400pt}{4.818pt}}
\put(463,68){\makebox(0,0){0.2}}
\put(463.0,857.0){\rule[-0.200pt]{0.400pt}{4.818pt}}
\put(585.0,113.0){\rule[-0.200pt]{0.400pt}{4.818pt}}
\put(585.0,857.0){\rule[-0.200pt]{0.400pt}{4.818pt}}
\put(706.0,113.0){\rule[-0.200pt]{0.400pt}{4.818pt}}
\put(706,68){\makebox(0,0){0.4}}
\put(706.0,857.0){\rule[-0.200pt]{0.400pt}{4.818pt}}
\put(828.0,113.0){\rule[-0.200pt]{0.400pt}{4.818pt}}
\put(828.0,857.0){\rule[-0.200pt]{0.400pt}{4.818pt}}
\put(950.0,113.0){\rule[-0.200pt]{0.400pt}{4.818pt}}
\put(950,68){\makebox(0,0){0.6}}
\put(950.0,857.0){\rule[-0.200pt]{0.400pt}{4.818pt}}
\put(1071.0,113.0){\rule[-0.200pt]{0.400pt}{4.818pt}}
\put(1071.0,857.0){\rule[-0.200pt]{0.400pt}{4.818pt}}
\put(1193.0,113.0){\rule[-0.200pt]{0.400pt}{4.818pt}}
\put(1193,68){\makebox(0,0){0.8}}
\put(1193.0,857.0){\rule[-0.200pt]{0.400pt}{4.818pt}}
\put(1314.0,113.0){\rule[-0.200pt]{0.400pt}{4.818pt}}
\put(1314.0,857.0){\rule[-0.200pt]{0.400pt}{4.818pt}}
\put(1436.0,113.0){\rule[-0.200pt]{0.400pt}{4.818pt}}
\put(1436,68){\makebox(0,0){1}}
\put(1436.0,857.0){\rule[-0.200pt]{0.400pt}{4.818pt}}
\put(220.0,113.0){\rule[-0.200pt]{292.934pt}{0.400pt}}
\put(1436.0,113.0){\rule[-0.200pt]{0.400pt}{184.048pt}}
\put(220.0,877.0){\rule[-0.200pt]{292.934pt}{0.400pt}}
\put(170,800){\makebox(0,0){$Re \, a$}}
\put(1350,65){\makebox(0,0){$\beta_{12}$}}
\put(220.0,113.0){\rule[-0.200pt]{0.400pt}{184.048pt}}

\put(244,332){\raisebox{-.8pt}{\makebox(0,0){\circle*{12}}}}
\put(244,571){\raisebox{-.8pt}{\makebox(0,0){\circle*{12}}}}
\put(244,571){\raisebox{-.8pt}{\makebox(0,0){\circle*{12}}}}
\put(244,810){\raisebox{-.8pt}{\makebox(0,0){\circle*{12}}}}
\put(269,332){\raisebox{-.8pt}{\makebox(0,0){\circle*{12}}}}
\put(269,572){\raisebox{-.8pt}{\makebox(0,0){\circle*{12}}}}
\put(269,572){\raisebox{-.8pt}{\makebox(0,0){\circle*{12}}}}
\put(269,810){\raisebox{-.8pt}{\makebox(0,0){\circle*{12}}}}
\put(293,331){\raisebox{-.8pt}{\makebox(0,0){\circle*{12}}}}
\put(293,572){\raisebox{-.8pt}{\makebox(0,0){\circle*{12}}}}
\put(293,572){\raisebox{-.8pt}{\makebox(0,0){\circle*{12}}}}
\put(293,810){\raisebox{-.8pt}{\makebox(0,0){\circle*{12}}}}
\put(317,330){\raisebox{-.8pt}{\makebox(0,0){\circle*{12}}}}
\put(317,573){\raisebox{-.8pt}{\makebox(0,0){\circle*{12}}}}
\put(317,573){\raisebox{-.8pt}{\makebox(0,0){\circle*{12}}}}
\put(317,810){\raisebox{-.8pt}{\makebox(0,0){\circle*{12}}}}
\put(342,328){\raisebox{-.8pt}{\makebox(0,0){\circle*{12}}}}
\put(342,574){\raisebox{-.8pt}{\makebox(0,0){\circle*{12}}}}
\put(342,574){\raisebox{-.8pt}{\makebox(0,0){\circle*{12}}}}
\put(342,810){\raisebox{-.8pt}{\makebox(0,0){\circle*{12}}}}
\put(366,326){\raisebox{-.8pt}{\makebox(0,0){\circle*{12}}}}
\put(366,574){\raisebox{-.8pt}{\makebox(0,0){\circle*{12}}}}
\put(366,574){\raisebox{-.8pt}{\makebox(0,0){\circle*{12}}}}
\put(366,810){\raisebox{-.8pt}{\makebox(0,0){\circle*{12}}}}
\put(390,324){\raisebox{-.8pt}{\makebox(0,0){\circle*{12}}}}
\put(390,575){\raisebox{-.8pt}{\makebox(0,0){\circle*{12}}}}
\put(390,575){\raisebox{-.8pt}{\makebox(0,0){\circle*{12}}}}
\put(390,810){\raisebox{-.8pt}{\makebox(0,0){\circle*{12}}}}
\put(415,322){\raisebox{-.8pt}{\makebox(0,0){\circle*{12}}}}
\put(415,577){\raisebox{-.8pt}{\makebox(0,0){\circle*{12}}}}
\put(415,577){\raisebox{-.8pt}{\makebox(0,0){\circle*{12}}}}
\put(415,811){\raisebox{-.8pt}{\makebox(0,0){\circle*{12}}}}
\put(439,319){\raisebox{-.8pt}{\makebox(0,0){\circle*{12}}}}
\put(439,578){\raisebox{-.8pt}{\makebox(0,0){\circle*{12}}}}
\put(439,578){\raisebox{-.8pt}{\makebox(0,0){\circle*{12}}}}
\put(439,811){\raisebox{-.8pt}{\makebox(0,0){\circle*{12}}}}
\put(463,316){\raisebox{-.8pt}{\makebox(0,0){\circle*{12}}}}
\put(463,579){\raisebox{-.8pt}{\makebox(0,0){\circle*{12}}}}
\put(463,579){\raisebox{-.8pt}{\makebox(0,0){\circle*{12}}}}
\put(463,811){\raisebox{-.8pt}{\makebox(0,0){\circle*{12}}}}
\put(488,313){\raisebox{-.8pt}{\makebox(0,0){\circle*{12}}}}
\put(488,581){\raisebox{-.8pt}{\makebox(0,0){\circle*{12}}}}
\put(488,581){\raisebox{-.8pt}{\makebox(0,0){\circle*{12}}}}
\put(488,811){\raisebox{-.8pt}{\makebox(0,0){\circle*{12}}}}
\put(512,309){\raisebox{-.8pt}{\makebox(0,0){\circle*{12}}}}
\put(512,582){\raisebox{-.8pt}{\makebox(0,0){\circle*{12}}}}
\put(512,582){\raisebox{-.8pt}{\makebox(0,0){\circle*{12}}}}
\put(512,811){\raisebox{-.8pt}{\makebox(0,0){\circle*{12}}}}
\put(536,306){\raisebox{-.8pt}{\makebox(0,0){\circle*{12}}}}
\put(536,584){\raisebox{-.8pt}{\makebox(0,0){\circle*{12}}}}
\put(536,584){\raisebox{-.8pt}{\makebox(0,0){\circle*{12}}}}
\put(536,812){\raisebox{-.8pt}{\makebox(0,0){\circle*{12}}}}
\put(560,302){\raisebox{-.8pt}{\makebox(0,0){\circle*{12}}}}
\put(560,586){\raisebox{-.8pt}{\makebox(0,0){\circle*{12}}}}
\put(560,586){\raisebox{-.8pt}{\makebox(0,0){\circle*{12}}}}
\put(560,812){\raisebox{-.8pt}{\makebox(0,0){\circle*{12}}}}
\put(585,298){\raisebox{-.8pt}{\makebox(0,0){\circle*{12}}}}
\put(585,588){\raisebox{-.8pt}{\makebox(0,0){\circle*{12}}}}
\put(585,588){\raisebox{-.8pt}{\makebox(0,0){\circle*{12}}}}
\put(585,812){\raisebox{-.8pt}{\makebox(0,0){\circle*{12}}}}
\put(609,294){\raisebox{-.8pt}{\makebox(0,0){\circle*{12}}}}
\put(609,590){\raisebox{-.8pt}{\makebox(0,0){\circle*{12}}}}
\put(609,590){\raisebox{-.8pt}{\makebox(0,0){\circle*{12}}}}
\put(609,812){\raisebox{-.8pt}{\makebox(0,0){\circle*{12}}}}
\put(633,289){\raisebox{-.8pt}{\makebox(0,0){\circle*{12}}}}
\put(633,592){\raisebox{-.8pt}{\makebox(0,0){\circle*{12}}}}
\put(633,592){\raisebox{-.8pt}{\makebox(0,0){\circle*{12}}}}
\put(633,813){\raisebox{-.8pt}{\makebox(0,0){\circle*{12}}}}
\put(658,285){\raisebox{-.8pt}{\makebox(0,0){\circle*{12}}}}
\put(658,594){\raisebox{-.8pt}{\makebox(0,0){\circle*{12}}}}
\put(658,594){\raisebox{-.8pt}{\makebox(0,0){\circle*{12}}}}
\put(658,813){\raisebox{-.8pt}{\makebox(0,0){\circle*{12}}}}
\put(682,280){\raisebox{-.8pt}{\makebox(0,0){\circle*{12}}}}
\put(682,596){\raisebox{-.8pt}{\makebox(0,0){\circle*{12}}}}
\put(682,596){\raisebox{-.8pt}{\makebox(0,0){\circle*{12}}}}
\put(682,813){\raisebox{-.8pt}{\makebox(0,0){\circle*{12}}}}
\put(706,275){\raisebox{-.8pt}{\makebox(0,0){\circle*{12}}}}
\put(706,598){\raisebox{-.8pt}{\makebox(0,0){\circle*{12}}}}
\put(706,598){\raisebox{-.8pt}{\makebox(0,0){\circle*{12}}}}
\put(706,814){\raisebox{-.8pt}{\makebox(0,0){\circle*{12}}}}
\put(731,270){\raisebox{-.8pt}{\makebox(0,0){\circle*{12}}}}
\put(731,600){\raisebox{-.8pt}{\makebox(0,0){\circle*{12}}}}
\put(731,600){\raisebox{-.8pt}{\makebox(0,0){\circle*{12}}}}
\put(731,814){\raisebox{-.8pt}{\makebox(0,0){\circle*{12}}}}
\put(755,265){\raisebox{-.8pt}{\makebox(0,0){\circle*{12}}}}
\put(755,603){\raisebox{-.8pt}{\makebox(0,0){\circle*{12}}}}
\put(755,603){\raisebox{-.8pt}{\makebox(0,0){\circle*{12}}}}
\put(755,814){\raisebox{-.8pt}{\makebox(0,0){\circle*{12}}}}
\put(779,260){\raisebox{-.8pt}{\makebox(0,0){\circle*{12}}}}
\put(779,605){\raisebox{-.8pt}{\makebox(0,0){\circle*{12}}}}
\put(779,605){\raisebox{-.8pt}{\makebox(0,0){\circle*{12}}}}
\put(779,815){\raisebox{-.8pt}{\makebox(0,0){\circle*{12}}}}
\put(804,255){\raisebox{-.8pt}{\makebox(0,0){\circle*{12}}}}
\put(804,607){\raisebox{-.8pt}{\makebox(0,0){\circle*{12}}}}
\put(804,607){\raisebox{-.8pt}{\makebox(0,0){\circle*{12}}}}
\put(804,815){\raisebox{-.8pt}{\makebox(0,0){\circle*{12}}}}
\put(828,250){\raisebox{-.8pt}{\makebox(0,0){\circle*{12}}}}
\put(828,610){\raisebox{-.8pt}{\makebox(0,0){\circle*{12}}}}
\put(828,610){\raisebox{-.8pt}{\makebox(0,0){\circle*{12}}}}
\put(828,816){\raisebox{-.8pt}{\makebox(0,0){\circle*{12}}}}
\put(852,245){\raisebox{-.8pt}{\makebox(0,0){\circle*{12}}}}
\put(852,612){\raisebox{-.8pt}{\makebox(0,0){\circle*{12}}}}
\put(852,612){\raisebox{-.8pt}{\makebox(0,0){\circle*{12}}}}
\put(852,816){\raisebox{-.8pt}{\makebox(0,0){\circle*{12}}}}
\put(877,240){\raisebox{-.8pt}{\makebox(0,0){\circle*{12}}}}
\put(877,614){\raisebox{-.8pt}{\makebox(0,0){\circle*{12}}}}
\put(877,614){\raisebox{-.8pt}{\makebox(0,0){\circle*{12}}}}
\put(877,817){\raisebox{-.8pt}{\makebox(0,0){\circle*{12}}}}
\put(901,234){\raisebox{-.8pt}{\makebox(0,0){\circle*{12}}}}
\put(901,617){\raisebox{-.8pt}{\makebox(0,0){\circle*{12}}}}
\put(901,617){\raisebox{-.8pt}{\makebox(0,0){\circle*{12}}}}
\put(901,817){\raisebox{-.8pt}{\makebox(0,0){\circle*{12}}}}
\put(925,229){\raisebox{-.8pt}{\makebox(0,0){\circle*{12}}}}
\put(925,619){\raisebox{-.8pt}{\makebox(0,0){\circle*{12}}}}
\put(925,619){\raisebox{-.8pt}{\makebox(0,0){\circle*{12}}}}
\put(925,818){\raisebox{-.8pt}{\makebox(0,0){\circle*{12}}}}
\put(950,223){\raisebox{-.8pt}{\makebox(0,0){\circle*{12}}}}
\put(950,622){\raisebox{-.8pt}{\makebox(0,0){\circle*{12}}}}
\put(950,622){\raisebox{-.8pt}{\makebox(0,0){\circle*{12}}}}
\put(950,819){\raisebox{-.8pt}{\makebox(0,0){\circle*{12}}}}
\put(974,218){\raisebox{-.8pt}{\makebox(0,0){\circle*{12}}}}
\put(974,624){\raisebox{-.8pt}{\makebox(0,0){\circle*{12}}}}
\put(974,624){\raisebox{-.8pt}{\makebox(0,0){\circle*{12}}}}
\put(974,820){\raisebox{-.8pt}{\makebox(0,0){\circle*{12}}}}
\put(998,212){\raisebox{-.8pt}{\makebox(0,0){\circle*{12}}}}
\put(998,627){\raisebox{-.8pt}{\makebox(0,0){\circle*{12}}}}
\put(998,627){\raisebox{-.8pt}{\makebox(0,0){\circle*{12}}}}
\put(998,820){\raisebox{-.8pt}{\makebox(0,0){\circle*{12}}}}
\put(1023,207){\raisebox{-.8pt}{\makebox(0,0){\circle*{12}}}}
\put(1023,629){\raisebox{-.8pt}{\makebox(0,0){\circle*{12}}}}
\put(1023,629){\raisebox{-.8pt}{\makebox(0,0){\circle*{12}}}}
\put(1023,821){\raisebox{-.8pt}{\makebox(0,0){\circle*{12}}}}
\put(1047,201){\raisebox{-.8pt}{\makebox(0,0){\circle*{12}}}}
\put(1047,632){\raisebox{-.8pt}{\makebox(0,0){\circle*{12}}}}
\put(1047,632){\raisebox{-.8pt}{\makebox(0,0){\circle*{12}}}}
\put(1047,822){\raisebox{-.8pt}{\makebox(0,0){\circle*{12}}}}
\put(1071,196){\raisebox{-.8pt}{\makebox(0,0){\circle*{12}}}}
\put(1071,634){\raisebox{-.8pt}{\makebox(0,0){\circle*{12}}}}
\put(1071,634){\raisebox{-.8pt}{\makebox(0,0){\circle*{12}}}}
\put(1071,822){\raisebox{-.8pt}{\makebox(0,0){\circle*{12}}}}
\put(1096,190){\raisebox{-.8pt}{\makebox(0,0){\circle*{12}}}}
\put(1096,636){\raisebox{-.8pt}{\makebox(0,0){\circle*{12}}}}
\put(1096,636){\raisebox{-.8pt}{\makebox(0,0){\circle*{12}}}}
\put(1096,823){\raisebox{-.8pt}{\makebox(0,0){\circle*{12}}}}
\put(1120,184){\raisebox{-.8pt}{\makebox(0,0){\circle*{12}}}}
\put(1120,639){\raisebox{-.8pt}{\makebox(0,0){\circle*{12}}}}
\put(1120,639){\raisebox{-.8pt}{\makebox(0,0){\circle*{12}}}}
\put(1120,824){\raisebox{-.8pt}{\makebox(0,0){\circle*{12}}}}
\put(1144,178){\raisebox{-.8pt}{\makebox(0,0){\circle*{12}}}}
\put(1144,641){\raisebox{-.8pt}{\makebox(0,0){\circle*{12}}}}
\put(1144,641){\raisebox{-.8pt}{\makebox(0,0){\circle*{12}}}}
\put(1144,825){\raisebox{-.8pt}{\makebox(0,0){\circle*{12}}}}
\put(1168,173){\raisebox{-.8pt}{\makebox(0,0){\circle*{12}}}}
\put(1168,644){\raisebox{-.8pt}{\makebox(0,0){\circle*{12}}}}
\put(1168,644){\raisebox{-.8pt}{\makebox(0,0){\circle*{12}}}}
\put(1168,826){\raisebox{-.8pt}{\makebox(0,0){\circle*{12}}}}
\put(1193,167){\raisebox{-.8pt}{\makebox(0,0){\circle*{12}}}}
\put(1193,646){\raisebox{-.8pt}{\makebox(0,0){\circle*{12}}}}
\put(1193,646){\raisebox{-.8pt}{\makebox(0,0){\circle*{12}}}}
\put(1193,827){\raisebox{-.8pt}{\makebox(0,0){\circle*{12}}}}
\put(1217,161){\raisebox{-.8pt}{\makebox(0,0){\circle*{12}}}}
\put(1217,648){\raisebox{-.8pt}{\makebox(0,0){\circle*{12}}}}
\put(1217,648){\raisebox{-.8pt}{\makebox(0,0){\circle*{12}}}}
\put(1217,828){\raisebox{-.8pt}{\makebox(0,0){\circle*{12}}}}
\put(1241,155){\raisebox{-.8pt}{\makebox(0,0){\circle*{12}}}}
\put(1241,651){\raisebox{-.8pt}{\makebox(0,0){\circle*{12}}}}
\put(1241,651){\raisebox{-.8pt}{\makebox(0,0){\circle*{12}}}}
\put(1241,829){\raisebox{-.8pt}{\makebox(0,0){\circle*{12}}}}
\put(1266,149){\raisebox{-.8pt}{\makebox(0,0){\circle*{12}}}}
\put(1266,653){\raisebox{-.8pt}{\makebox(0,0){\circle*{12}}}}
\put(1266,653){\raisebox{-.8pt}{\makebox(0,0){\circle*{12}}}}
\put(1266,830){\raisebox{-.8pt}{\makebox(0,0){\circle*{12}}}}
\put(1290,144){\raisebox{-.8pt}{\makebox(0,0){\circle*{12}}}}
\put(1290,655){\raisebox{-.8pt}{\makebox(0,0){\circle*{12}}}}
\put(1290,655){\raisebox{-.8pt}{\makebox(0,0){\circle*{12}}}}
\put(1290,831){\raisebox{-.8pt}{\makebox(0,0){\circle*{12}}}}
\put(1314,138){\raisebox{-.8pt}{\makebox(0,0){\circle*{12}}}}
\put(1314,658){\raisebox{-.8pt}{\makebox(0,0){\circle*{12}}}}
\put(1314,658){\raisebox{-.8pt}{\makebox(0,0){\circle*{12}}}}
\put(1314,833){\raisebox{-.8pt}{\makebox(0,0){\circle*{12}}}}
\put(1339,132){\raisebox{-.8pt}{\makebox(0,0){\circle*{12}}}}
\put(1339,660){\raisebox{-.8pt}{\makebox(0,0){\circle*{12}}}}
\put(1339,660){\raisebox{-.8pt}{\makebox(0,0){\circle*{12}}}}
\put(1339,834){\raisebox{-.8pt}{\makebox(0,0){\circle*{12}}}}
\put(1363,126){\raisebox{-.8pt}{\makebox(0,0){\circle*{12}}}}
\put(1363,662){\raisebox{-.8pt}{\makebox(0,0){\circle*{12}}}}
\put(1363,662){\raisebox{-.8pt}{\makebox(0,0){\circle*{12}}}}
\put(1363,835){\raisebox{-.8pt}{\makebox(0,0){\circle*{12}}}}
\put(1387,120){\raisebox{-.8pt}{\makebox(0,0){\circle*{12}}}}
\put(1387,664){\raisebox{-.8pt}{\makebox(0,0){\circle*{12}}}}
\put(1387,664){\raisebox{-.8pt}{\makebox(0,0){\circle*{12}}}}
\put(1387,837){\raisebox{-.8pt}{\makebox(0,0){\circle*{12}}}}
\put(1412,114){\raisebox{-.8pt}{\makebox(0,0){\circle*{12}}}}
\put(1412,666){\raisebox{-.8pt}{\makebox(0,0){\circle*{12}}}}
\put(1412,666){\raisebox{-.8pt}{\makebox(0,0){\circle*{12}}}}
\put(1412,838){\raisebox{-.8pt}{\makebox(0,0){\circle*{12}}}}
\end{picture}
%%%%%%%%%%%%%%
\caption{Real parts of eigenvalues as functions of non diagonal friction. 
Underdamped regime, $\omega=0.6>\beta/2=0.5.$ Since two eigenvalues
are complex conjugate, three curves only appear.}
\end{figure}

Hence, when $t \rightarrow \infty,$ only the ``$a_1$ mode'' survives. The 
leading terms in Eqs.(\ref{q1}) and (\ref{sigma1}) for long times are,
\be
<q_1(t)>=\theta_{11} \, X_{10} \, e^{a_1 t}, \ \ \ \ \ \ \sigma^2_{q1}(t)=
\frac{\theta_{11}^2 \,  e^{2a_1t} \, (\alpha \alpha^T)_{11}} {2a_1}.
\ee
The result for $t \rightarrow \infty$ reads,
\be
P(q_{10},q_{20},p_{10},p_{20})\ \rightarrow\ \frac12\, {\rm erfc} 
\left(-\frac{X_{10}}{\sqrt{2 \bar A_{11}}}\right), \ \ \ \ \ \ 
\bar A_{11}= (\alpha \alpha^T)_{11} / (2a_1).
\ee
It is similar to the result obtained in 1-D, provided we choose the proper 
coordinate,
\be
X_1=\omega_1^2 q_1 + a_1 p_1 - \beta_{12} 
\frac{a_1}{\omega_2^2+a_1^2+\beta_2 a_1}(-\omega_2^2 q_2+a_1 p_2).
\label{X1}\ee

The condition to have half of the particles to overpass the barrier is then 
$X_{10}=0$. When the two degrees of freedom are uncorrelated ($\beta_{12}=0$),
one gets the same condition as in one dimension, Eq.(\ref{extra}). But in 
general, it is not possible to simply express it in terms of the initial 
kinetic energy because it also depends on the orientation of the initial 
velocity in the potential landscape. For the same reasons, the diffusion 
time is not easy to evaluate either. Note that in the previous graphs, 
(Figs. 1 \& 2), $a_1$ is almost constant 
and could be approximated by the value given in Eqs.(\ref{eigen}). 

Then we can evaluate $\bar A_{11}$, 
\be
\bar A_{11}=\frac{T}{a_1}(\beta_1 (\theta^{-1})_{13}^2 + \beta_2
(\theta^{-1})_{14}^2 +2 \beta_{12}
(\theta^{-1})_{13}(\theta^{-1})_{14}),
\ee
provided the diagonalisation of the drift matrix can be done. Note
that only $(\theta^{-1})_{13}$ and $(\theta^{-1})_{14}$ occurs, which
are coupling $X_1$ with the velocity coordinates $p_1$ and
$p_2$. From Eq. (\ref{X1}), one gets,
\beqn
\bar A_{11}=Ta_1\left(\beta_1-\beta_{12}^2 a_1^2 \frac{2\omega_2^2 + 2
a_1^2 +\beta_2 a_1}{(\omega_2^2+a_1^2+\beta_2a_1)^2}\right).
\eeqn
Again, when the two degrees of freedom are uncorrelated ($\beta_{12}=0$),
one gets the same probability as in one dimension, Eqs.(\ref{passed},\ref{lt}).

%%%%%%%%%%%%%%%%%%%%%%%%%%%% Conclusion

\section{Conclusion}

In this paper, we showed a general scheme to solve multi-dimensional 
Langevin equations near a saddle point. In one dimension, the solution is 
very simple and the diffusion condition, time and probability can be 
easily expressed. This means that in the case of a simple one dimension 
model for heavy-ion fusion, one can analytically estimate the extra-push 
energy, the fusion time and its probability.

For higher dimensions, the stochastic dynamics is again easily solved, but a 
difficulty remains, namely the explicit diagonalization of the drift matrix. 
In such conditions, a general analytic solution cannot be written, but the 
main features we found could be easily applied to very specific physical 
problems. In particular, we showed the difference between the dominant 
degree of freedom and the damped ones. The possible occurrence of complex
eigenvalues corresponds to residual oscillations in subdominant degrees.

In every case, the Gaussian solution to the problem contradicts the naive,
intuitive expectation of a distribution with two peaks moving apart from
the fusion barrier. In fact, the way over the barrier simply results from a 
competition between the drift of the center of the Gaussian and its spreading.

For exotic noises leading to anomalous diffusion, such as L\'evy flights, 
the Langevin equation can be analytically solved for a one dimension 
overdamped motion where the Langevin equation reduces to a Smoluchowski one. 
In that case, the solution is given in Ref. \cite{Jes}.

%\vspace{2cm}
{\bf{\it Acknowledgments}} We are indebted to D. Kusnezov for an interesting 
discussion on anomalous diffusion and L\'evy flights in particular. 
Two of us (D.B. and B.G.G) thank the Yukawa Institute for Theoretical Physics 
for its warm hospitality, during visits where part of this work was done.

%%%%%%%%%%%%%%%%%%%%%%%%%%%% Bibliography

\ 

\ 

\centerline{{\bf Appendix A}}

\ 

In order to avoid possible confusions brought by the three successive
changes of representation expressed by the matrices $M^{-1/2},$ $U$
and $\theta,$ we prove again the well-known fluctuation-dissipation theorem. 
Our precise purpose is to parameterize the random force $R,$ see 
Eq.(\ref{canoni}), in terms of dimensionless, independent random numbers 
$\nu_i,$ normalized to unity, see Eq.(\ref{norma}). The starting point 
is a simplified form of Eq.(\ref{brute}), namely a situation where just 
the friction and the random force are present,
\be
M \ddot Z + {\cal G} \dot Z =  {\cal F}(t) 
\, .
\ee
This leads at once to a simplified form of Eq.(\ref{demi}),
\be
M^{1/2} \ddot Z + M^{-1/2}{\cal G} M^{-1/2} M^{1/2}\dot Z = 
M^{-1/2} {\cal F}(t) 
\, .
\label{demisimpl}
\ee
It is convenient at this stage to diagonalize the real, symmetric, positive
definite matrix ${\cal B} \equiv M^{-1/2}{\cal G}M^{-1/2},$ and obtain its
representation in terms of its eigenvectors $|i>$ and eigenvalues $\lambda_i,$
\be
{\cal B}=\sum_{i=1}^n |i>\lambda_i<i|.
\ee
This decouples Eq.(\ref{demisimpl}) into independent modes, relaxing 
separately towards thermal equilibrium for large times,
\be
<i|M^{1/2} \ddot Z >=-\lambda_i<i|M^{1/2} \dot Z>+<i|M^{-1/2}{\cal F}(t)>,
\ee
namely
\be
<i|M^{1/2} \dot Z >\ \rightarrow_{t \rightarrow + \infty} 
\int_0^t d\tau \, e^{-\lambda_i(t-\tau)} <i|M^{-1/2}{\cal F}(\tau)> 
\, .
\label{relax}
\ee
The left-hand side of this limit, Eq.(\ref{relax}), can now be squared 
as a kinetic energy and then ensemble averaged to be equated with the 
Boltzmann energy $T=\overline {(<i|M^{1/2} \dot Z >)^2} \, .$ 
(We set the Boltzmann constant as a unit, $k=1,$ as usual.) 
The right-hand side of Eq.(\ref{relax}), in turn, can be 
constrained by the following ansatz for ensemble averaging,
\be
\overline {<i|M^{-1/2}{\cal F}(\tau)> <i|M^{-1/2}{\cal F}(\tau')>} =  \rho_i
\, \delta(\tau-\tau'),
\ee
where $\rho_i$ is an unknown normalization. The square of the
right-hand side of Eq.(\ref{relax}) thus becomes,
\be
\rho_i \int_0^{t\rightarrow\infty} d\tau\, e^{-2\lambda_i(t-\tau)}=
\frac{\rho_i}{2 \lambda_i},
\ee
hence the normalization result, $\rho_i=2 T\lambda_i.$
Let $\Lambda$ be the diagonal matrix defined by the square roots 
$(2T\lambda_i)^{1/2}$ of such normalizations $\rho_i.$ Let ${\cal U}$
be that orthogonal matrix which  lists the eigenvectors $|i>$ as
columns. Let $\nu \equiv \{\nu_1,...,\nu_n\}$ be a column vector of 
normalized, Gaussian, independent random numbers. Since our 
fluctuation-dissipation theorem states that 
$M^{-1/2}{\cal F}=\sum_{i=1}^n |i> (2T\lambda_i)^{1/2}\,\nu_i
={\cal U}\, \Lambda\, \nu,$ 
the stochastic force $R$ present in Eq.(\ref{canoni}) 
reads $R=\Gamma\,\nu,$ with $\Gamma=U^{-1}\,{\cal U}\,\Lambda.$ 
This explains why $\Gamma$ is not expected to be symmetric.Eventually,
the fluctuation-dissipation can be written in a more classical way,
\be
\Gamma\Gamma^T=2T\beta,
\ee 
where $\beta=U^{-1} M^{-1/2}{\cal G}M^{-1/2}U$ is the reduced friction matrix.  

\ 

\centerline{{\bf Appendix B}}
\newcommand{\dt}{{\delta\tau}}
\ 

The solution of multi-dimensional Langevin equations is shown here for 
Markovian noises.

Using the fact that the Euler type variables, $(x_1,\ldots,x_{2n})$, have 
the same statistical properties as those of Eq.(\ref{Eulb}), we find,
\beqn
W(x_1,...,x_{2n},t;X_{10},...,X_{2n0}) &=&
<\delta[x_1-x_1(t)]...\delta[x_{2n}-x_{2n}(t)]>,\\
&=& \int \frac{dk_1}{2\pi} ... \frac{dk_{2n}}{2\pi} 
<\exp{\left(i[k_1,..,k_{2n}]  
\left[\matrix{x_1 - x_1(t) \cr \dot{.} \cr x_{2n}- x_{2n}(t) }\right]\right)}>,
\eeqn
where the $x_i(t)$'s are shown in Eq.(\ref{Eula}). The time integral will be 
discretized, with $\dt=t/L,$
\be
\left[\matrix{x_1(t)\cr\dot{.}\cr x_{2n}(t)}\right] = \int_0^t d\tau\, e^{-\tau
{\rm D}}\  \alpha  
\left[\matrix{\nu_1(\tau) \cr\dot{.}\cr\nu_n(\tau)}\right]
=\lim_{L \rightarrow \infty}\sum_{\ell=1}^L  e^{-(\ell-1/2)\dt
{\rm D}}\  \alpha  
\left[\matrix{\nu_1(\dt) \cr\dot{.}\cr\nu_n(\dt)}\right].
\ee
Here, ${\rm D}$ is the diagonal drift matrix, see Eq.(\ref{DDM}) and 
$\nu_j(\delta\tau)=\int_{(\ell-1)\dt}^{\ell\, \dt}\, d\tau\, \nu_j(\tau)$ 
are the results of random walks in the force space during the time interval 
$\dt$. Then the distribution function can be evaluated in its Fourier space,
using the statistical properties of such random walks,
\be
W(x_1,...,x_{2n},t;X_{10},...,X_{(2n)0}) = \int \frac{dk_1}{2\pi} ...
\frac{dk_{2n}}{2\pi} \exp{\left(i\, [k_1,..,k_{2n}]  
\left[\matrix{x_1 \cr \dot{.} \cr x_{2n}}\right]\right)}\, p(k_1,...,k_{2n},t),
\ee
with 
\be
p(k_1,...,k_{2n},t)=\lim_{L \rightarrow \infty}\prod_{\ell=1}^L
<\exp{\left(-i [k_1,..,k_{2n}] e^{-(\ell-1/2)\dt {\cal A}}\ \alpha  
\left[\matrix{\nu_1(\dt) \cr\dot{.}\cr\nu_n(\dt)}\right]\right)}>.
\label{charac}\ee
We have used the fact that the random numbers are Markovian to say that the 
average of a product is the product of the averages of its factors.

In the case of a Gaussian noise, the distribution function of the random 
numbers reads,
\be
p(\nu_1,..,\nu_n)= \frac1{\sqrt{(2\pi)^n}} 
\exp{\left(-\frac12\, [\nu_1,..,\nu_n]
\left[\matrix{\nu_1\cr \dot{.} \cr \nu_n}\right]\right)},
\ee 
and, due to their Markovian properties, their integration over a time step 
$\dt$ leads to
\be
p(\nu_1(\dt),..,\nu_n(\dt))= \frac1{\sqrt{(2\pi\dt)^n}}\,
\exp{\left(-\frac1{2\dt}\, [\nu_1(\dt),..,\nu_n(\dt)]\,
\left[\matrix{\nu_1(\dt)\cr \dot{.} \cr \nu_n(\dt)}\right]\right)}.
\ee
Therefore,
\be
p(k_1,...,k_{2n},t)= \lim_{L \rightarrow \infty} \prod_{\ell=1}^L
\exp{\left(-\frac{\dt}2\, [k_1,..,k_{2n}]\, e^{-(\ell-1/2)\dt D}\ \alpha
 \alpha^T\  e^{-(\ell-1/2)\dt D}\, \left[\matrix{k_1\cr \dot{.} 
\cr k_{2n}}\right]\right)}.
\ee
Reintroducing the time integral, one obtains,
\be
p(k_1,...,k_{2n},t)= \exp{\left(-\frac12\, [k_1,..,k_{2n}]  
A\, \left[\matrix{k_1\cr \dot{.} \cr k_{2n}}\right]\right)},
\ee
where the matrix $A$ is defined by Eq.(\ref{A}). Inverting the Fourier 
transform, upon $p(k_1,...,k_{2n},t)$, one eventually gets, 
\be
W(x_1,...,x_{2n},t;X_{10},...,X_{(2n)0}) = \frac1{(2\pi)^n} 
\frac1{\sqrt{\det A}} 
\exp\left(-\frac12 [x_1,..,x_{2n}] A^{-1} \left[\matrix{x_1 \cr \dot{.} 
\cr x_{2n}}\right] \right).
\ee

For L\'evy flights, we will restrict ourselves to a one dimension problem. The
noise is defined by its characteristic function $p(k)$ in the Fourier space, 
\be
p(k)=\int d\nu\, e^{-ik\nu} p(\nu) = \exp{[-\Delta |k|^{\mu}]},
\ee
and is the source of an anomalous behavior characterized by a mean square 
displacement of the form $<\sigma_{q}(t)>\propto 2 \Delta t^{\gamma}$ 
\cite{Bou}. As the
fluctuation-dissipation theorem is not satisfied any more, we keep a 
generalized diffusion coefficient, $\Delta$. For the special case $\mu=2$, the
noise is Gaussian. Similarly to the Gaussian case, the integration over a 
time step $\dt$ leads to,
\be
p'(k)=\int d\nu(\dt)\, e^{-ik\nu(\dt)}\, p(\nu(\dt)) = 
\exp{[-\Delta (\dt)^{1-\mu} |k|^{\mu}]},
\ee
after renormalization \cite{Fog}. Therefore, the average value in the r.h.s. 
of Eq.(\ref{charac}) can be calculated,
\be
<\exp{[-i\, \dt\, \nu_i(\dt)\, (k_1 e^{-a(\ell-1/2)\dt} + k_2  
e^{-b(\ell-1/2)\dt})]}> = 
\exp{[- \Delta \dt\, (k_1 e^{-a(\ell-1/2)\dt} + k_2  
e^{-b(\ell-1/2)\dt})^{\mu}]}.
\ee
Reintroducing the time integral, one finally gets,
\be
p(k_1,k_2,t)=\exp{\left[-\Delta \int_0^t (k_1 e^{-a\tau} + k_2 
e^{-b\tau})^{\mu} \,d\tau\right]}.
\ee
For $\mu=2$, i.e. in the Gaussian case, the time integral can be 
evaluated analytically, but such is not the case for any value of $\mu$.
The only favorable situation occurs for a one dimension Smoluchowski equation 
where,
\be
p(k_1,t)=\exp{\left(-\Delta \int_0^t k_1^{\mu}\, e^{-a\mu\tau} d\tau\right)},
\ee
see Ref. \cite{Jes}. This is why we restrict our study to classical 
Gaussian noises.

\end{document}